\documentclass[authoryear,preprint,12pt]{elsarticle}

\usepackage[margin=1in]{geometry}
\usepackage{epsfig}
\usepackage{hyperref}
\usepackage{amsmath,amsfonts,amssymb,amscd,amsthm,xspace}
\usepackage{color}
\usepackage{rotating}
\usepackage{soul} 	
\usepackage{appendix}

\journal{}

\begin{document}
\begin{frontmatter}
\hypersetup{
    linkcolor=red,          
    citecolor=green,        
    filecolor=magenta,      
    urlcolor=cyan           
}

\title{Mathematical model of galactose regulation and metabolic consumption in yeast}

\author[math]{Tina M.~Mitre}
\author[math,phgy,nldc]{Michael C.~Mackey}
\author[math,phgy,nldc]{Anmar~Khadra\corref{cor1}}
\ead{anmar.khadra@mcgill.ca}

\cortext[cor1]{Corresponding author}

\address[math]{Department of Mathematics, McGill University, Montreal, Quebec, Canada}
\address[phgy]{Department of Physiology, McGill University, Montreal, Quebec, Canada}
\address[nldc]{Centre for Nonlinear Dynamics, Montreal, Quebec, Canada}


\begin{abstract}
The galactose network is a complex system responsible for galactose metabolism. It has been extensively studied experimentally and mathematically at the unicellular level to broaden our understanding of its regulatory mechanisms at higher order species. Although the key molecular players involved in the metabolic and regulatory processes underlying this system have been known for decades, their interactions and chemical kinetics remain incompletely understood. Mathematical models can provide an alternative method to study the dynamics of this network from a quantitative and a qualitative perspective. Here, we employ such approaches to unravel the main properties of the galactose network, including equilibrium binary and temporal responses, as a way to decipher its adaptation to actively-changing inputs. We combine the two main components of the network; namely, the genetic branch which allows for bistable responses, and a metabolic branch, encompassing the relevant metabolic processes and glucose repressive reactions. We use both computational tools to estimate model parameters based on published experimental data, as well as bifurcation analysis to decipher the properties of the system in various parameter regimes. Our model analysis reveals that the interplay between the inducer (galactose) and the repressor (glucose) creates the bistability regime which dictates the temporal responses of the system. Based on the same bifurcation techniques, we can also explain why the system is robust to genetic mutations and molecular instabilities. These findings may provide experimentalists with a theoretical framework upon which they can determine how the galactose network functions under various conditions.
\end{abstract}
\end{frontmatter}

\section{Introduction} 
\label{sec:intro}
\noindent
Experimental studies of genetic regulatory networks in unicellular organisms are based, at least in part, on the premise that the main regulatory mechanisms are conserved across species and thus basic biochemical constituents  are utilized in the same way, regardless of the complexity of the organism. The galactose network is a typical example of such networks that has been extensively studied in the budding yeast \textit{Saccharomyces cerevisiae}. It is comprised of metabolic reactions coupled to a set of genetic regulatory processes and glucose-repressed proteins involved in galactose utilization. This network is typically activated when galactose, a monosaccharide found in dairy and vegetables, becomes the only available energy source, triggering a cascade of intracellular processes that can be repressed in the presence of glucose.

The protein machinery of the galactose network comprises $\sim 5 \%$ of the total cellular mass \citep{bhatbook}. Because of this high protein load, galactose is energetically more expensive to use than glucose. As a result, cells use glucose as a transcriptional repressor of the proteins of the GAL network (Gal proteins) when both monosaccharides (glucose and galactose) are present. Although not an essential nutrient, galactose is a crucial moiety in the cellular membrane glycoproteins \citep{dejongh2008}. Genetic mutations in the amino-acid sequence of the galactose metabolic enzymes lead to the accumulation of galactitol, an  alcohol-form of galactose, which can be lethal if no food restrictions are applied. Such genetic disorders fall under the pathological condition ``galactosemia'', which currently affects 1 in 600,000 children \citep{murphy1999}. Secondary effects of this disease include cataracts and neuronal degenerative disorders \citep{timson2006}. For these health reasons, it is imperative to obtain a thorough understanding of the gene regulatory network and the associated metabolic pathways underlying galactose regulation.

Interest in the Gal proteins appeared first in the 1940's with work by \citet{kosterlitz1943} that focused on galactose fermentation and the related metabolic system in budding yeast. Mathematical modeling of this system, however, started appearing in the late 1990's with work by \citet{venkatesh1999}, focusing on the regulatory GAL network that consisted of three feedback loops. Since then, several groups have worked on variations of this study by developing models with different degrees of complexity to understand previous experimental results \citep{atauri2004,atauri2005,ramsey2006,acar2005,acar2010,apostumackey,venturelli2012}.

For example, de Atauri and colleagues have published four papers on galactose metabolic and gene regulation, with an emphasis on transcriptional noise. From a general perspective, \citet{atauri2004} argued for the advantages of using dynamic models to describe such biological systems. These models were used to further analyze and explain certain aspects associated with GAL network repression, the switch-like phenomenon and the tight control that keeps the galactose metabolite Galactose-1-Phosphate (Gal1P) below an (unknown) toxicity threshold. Having shown possible links to the genetic disease galactosemia in experimental settings, Gal1P was the main subject of interest in \citet{atauri2005}. The study showed that the metabolic concentrations are kept at low intracellular levels, whereas the control machinery works to attenuate high-frequency noise. The control of these concentrations at a low stable level was further analyzed in \citet{ramsey2006}, using a combination of determining induction curves for wild-type and GAL3-GAL80 mutant strains from flow cytometry data, model fitting to the experimental results and statistical analysis of stochastic simulations. The authors demonstrated that positive and negative regulatory feedback loops, mediated through the Gal3 and the Gal80 proteins respectively, are necessary to avoid large intracellular variations and long initial transients in the induction phase of network. Based on an extension of their mathematical modeling and a novel experimental approach, \citet{bennett2008} then characterized the effects of oscillations in glucose on the network. By analyzing transcription fluorescence levels, they concluded that the network behaves as a low pass filter but did not offer a complete analysis of the properties of the system.

In 2005, Acar and colleagues studied the effect of the different feedback loops through mutations in the GAL mRNA strains, from both experimental and mathematical perspectives \citep{acar2005}. They showed that different induction curves with a bistable behaviour appear over a limited range of galactose concentrations for all yeast strains they examined except for the GAL3 mutant, which encodes for a regulatory protein of the network. The work in \citet{acar2010}, on the other hand, focused on describing gene networks and the importance of feedback mechanisms on the number of gene copies in a cell. More specifically, this latter work investigated the induction of the GAL network and how it can be hindered by removing or diminishing various feedback loops in the system. Through computational and experimental observations, the study showed the existence of a 1-to-1 stoichiometry between Gal3 and Gal80 proteins in the gene network.

The bistability property was further analyzed in more recent work. For example, in a study by \citet{venturelli2012}, the induction of the GAL network published in \citet{acar2005} was reassessed experimentally using flow cytometry methods and qualitatively using mathematical modeling. The study presented a mathematical model that maintained the bistable behaviour seen in  \citep{acar2005}, but it was developed based on simple feedback mechanisms in which GAL3, GAL80 and GAL1 transcription rates depended on the Gal4 protein in a Michaelis-Menten fashion and Gal3 and Gal1 received a constant input rate upon galactose administration to the cell. Interestingly, the model included a recently-discovered positive feedback loop of the galactose network involving the Gal1 protein (Gal1p) \citep{abramczyk2012}. The experimental and modeling results of \citet{venturelli2012} demonstrated that bistability occurs due to pathways involving both Gal3 (Gal3p) and Gal1 proteins, contradicting previous experimental results by \citep{acar2005}. Moreover, it concluded that the interplay between these two positive feedback loops increases the bistability range of the system and that connections of this kind can be beneficial in nature as it may induce a faster response time to abrupt environmental changes than a single positive loop. In  \citet{apostumackey}, the exact sequence of reactions occurring at the promoter level of GAL genes was analyzed mathematically to determine how bistability is affected by model variations involving Gal3p (activated by galactose) and to pinpoint which variation produces results most consistent with those seen in  \citet{acar2010}. Their results showed that the GAL regulon is induced at the promoter level by Gal3p activated dimers through a non-dissociation sequential model.

Our interest in the galactose network is two-fold: first, the GAL regulon is one of the operons found in unicellular organisms that may share several important dynamic properties with other regulons, such as the lactose regulatory network; second, mathematical modeling of such processes can allow us to understand how various extracellular perturbations affect the genetic system, its memory and filtering capacities.

Our goal is to characterize the GAL network using a mathematical modeling approach which includes the ``genetic" model of \citet{apostumackey}, coupled to four different glucose-repression events and a simplified metabolic pathway. The model takes into account the major processes responsible for the determination of intracellular galactose dynamics: Gal3 and Gal1 activation, galactose transport through the Gal2 permease, phosphorylation by Gal1 kinase and dilution due to cell growth. The model reveals that bistability not only persists in the full GAL metabolic-gene network, even after the addition of other feedback loops involved in the metabolic pathways, but is also dynamically robust (i.e., exhibited over a wide range of parameters). This numerical result suggests that the organism is adaptable to various conditions. The model is then examined to determine its sensitivity to different concentrations of the repressor (glucose) and its adaptability to a repressive oscillatory signal at different 
frequencies.

The paper is organized as follows:  Section \ref{model} details the complete development of the combined genetic and metabolic model for galactose regulation, taking into account the underlying biochemistry and gene regulatory aspects. Section \ref{results} presents the analytical and numerical results, particularly how the full galactose network exhibits bistability and how the model adapts to oscillatory perturbations in extracellular glucose concentrations. The paper ends with the discussion and conclusions in Section \ref{discussion}, where we detail a number of experimentally testable predictions deduced from our results and suggest experimental steps on how to verify them.
\section{Mathematical modeling of the GAL regulon and the Leloir pathway} \label{model}
\noindent
When discussing the cellular processes affected by galactose, we typically focus on two main branches: (i) the metabolic branch, or the Leloir pathway, responsible of converting galactose into other forms suitable for energy consumption; and (ii) the genetic branch that consists of regulatory processes happening on a slower time scale than the ones occurring in the previously mentioned branch. The two main components of the Leloir pathway are the Gal2p permease, the main transporter of galactose, and the Gal1p kinase, which transforms intracellular galactose into a phosphorylated form. For the genetic branch, the main molecular players are the mRNAs, exhibiting high transcription levels upon galactose induction (i.e., GAL3, GAL80, GAL2 and GAL1 mRNAs) along with the Gal80, Gal3 and Gal1 proteins, acting as regulators at the promoter level. Among these proteins, Gal80p down-regulates the expression of all Gal proteins, whereas Gal3p and Gal1p activate the network. This means that galactose activates several 
feedback loops once transported to the cytosol. The other important sugar in yeast is raffinose, a trisaccharide composed of fructose, glucose and galactose. With respect to the galactose network, raffinose is a non-inducible, non-repressible medium, similar to glycerol \citep{stockwell2015}. In its presence, a leakage is observed in the mRNA levels of GAL80 and GAL3, leading to a 3 to 5 fold increase in their expression levels \citep{giniger1985}.

In the following model development, proteins of the galactose network are denoted by small letters (e.g. Gal1p and Gal2p), whereas capital letters are used for genes (e.g. GAL3 and GAL2). GAL3, GAL80, GAL2 and GAL1 mRNA expression levels are denoted by $M_3$,  $M_{80}$,  $M_2$ and $M_1$, respectively, whereas their protein concentrations are denoted by $G_3,$  $G_{80}$, $G_2$ and $G_1$, respectively. Throughout our analysis, we will assume that protein translation is directly proportional to the expression level of mRNA produced.

\subsection{GAL regulon} \label{galreg}
\begin{figure}[!htb]
\centering
\includegraphics[width=12cm]{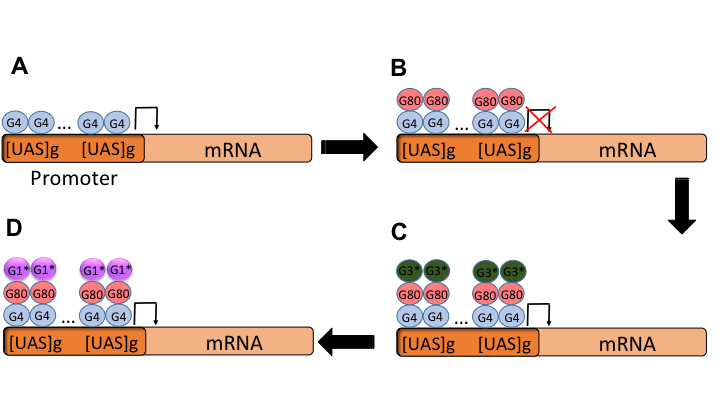}
\caption[Schematic illustration of the GAL regulon]{Schematic illustration of the GAL regulon and the  effects of the proteins on their own
transcription: (A) Gal4p dimers ($[G_4:G_4]$) bind to
the upstream activating sequence ($[UAS]_g$) on the promoter region of GAL mRNAs
and induce transcription. (B) When galactose is absent from the external medium, Gal80p dimers
($[G_{80}:G_{80}]$) bind to Gal4p dimers and repress
the aforementioned process. (C) In the presence of galactose, Gal3 proteins are
activated ($G_3^*$) and acquire a high affinity binding to the $[G_4:G_4]:[G_{80}:G_{80}]$ complex. The formation of this new multi-complex reactivates the network, inducing all Gal proteins and giving rise to a
positive feedback loop. (D) In the long term, experimental evidence
suggests that Gal1p dimers ($[G_1^*:G_1^*]$) replace Gal3p dimers
($[G_3^*:G_3^*]$) which leads to further and ``more
efficient" transcriptional induction \citep{abramczyk2012,zack2007}. The
repeated dots ($\cdots$) in each panel represent the series of binding reactions undertaken by each mRNA species considered.}
\label{fig:galregulon}
\end{figure}

\noindent
The gene regulation part of our model combines assumptions from \citet{apostumackey} with recent experimental results on the existence of a positive feedback loop mediated by Gal1p \citep{abramczyk2012}. A similar model including this additional pathway has been published in \citet{venturelli2012}, but complete chemical reactions and dimerization processes were not considered. The kinetic reactions pertaining to the gene regulatory network are shown schematically in Fig. \ref{fig:galregulon}.

As indicated by panel A of Fig. \ref{fig:galregulon}, Gal4p dimers ($G_{4d}$) have a high affinity for regions of the GAL promoter known as upstream activating sequences ($[UAS]_g$), which are 17 base-pair sequences. Depending on the respective mRNA species, these sequences can occur more than once. It has been shown experimentally that for the GAL3 and GAL80 genes, there is a single $[UAS]_g$, while for the other GAL mRNAs, there are up to 5 sequences \citep{ideker2001,sellickreece2008,atauri2004}. The $G_{4d}:[UAS]_g$ complex has a high affinity for the Gal80 dimer ($G_{80d}$) when raffinose is present (panel B). This dimer acts as an inhibitor and therefore creates a negative feedback on its own transcription. The second row of Fig. \ref{fig:galregulon} displays the reactions that take place in the presence of galactose. First, transcription is induced because Gal3p dimers, 
activated by 
galactose, remove the inhibition exerted by Gal80p dimers (panel C), generating the tripartite complex $[G_{4d}:G_{80d}:G_{3d}]$ in high proportion in the nucleus during the first 10 minutes of galactose induction \citep{abramczyk2012}. This is then followed by the substitution of Gal3 by Gal1p dimers to form the new tripartite complex $[G_{4d}:G_{80d}:G_{1d}]$ (panel D). Moreover, it was previously mentioned that these two proteins bind to galactose and ATP, that they share 70\% of base pairs and 90\% of their amino-acid sequence \citep{bhatbook}. This high degree of homology is consistent with their shared role at the promoter level.

\begin{sidewaystable}
\begin{center}
\begin{tabular}   { c  c  l  l  c }  \hline \vspace{-6pt}\\
\textbf{Energy}&\textbf{Effect on the}&\textbf{Description}&\textbf{Process}&\textbf{Figure}\\
\textbf{Source}&\textbf{GAL Network}& & & \vspace{6pt}\\ \hline
Glucose&Repression& $[UAS]_g \rightarrow [UAS]_g:M_{1p}$	& Mig1p protein binds to the upstream activating& \\
 & 	& 											&sequence ($[UAS]_{g}$) on the GAL genes and &\\
 										& 	&	&refrains the RNA polymerase from transcription.&\\ \hline
 Raffinose &Induction	& $[UAS]_g \rightarrow D_1$ 		&G4 dimers ($[G_{4d}]$) bind to $[UAS]_{g}$ and allow&\ref{fig:galregulon}(A)\\
 		& Leakage 	&  							&transcription to occur.&\\ \hline
Raffinose, &Induction 	& $G_{80}+G_{80} \xrightarrow{K_{D,80}} G_{80,d}$& $G_{80}$ dimers ($[G_{80d}$]) bind to the previously formed &\\
Galactose  		& Leakage	&  $D_1 + G_{80,d} 
\xrightarrow{K_{B,80}} D_{2}$										& complex. They inhibit Gal protein transcription, with&\ref{fig:galregulon}(B)\\
&	& &some leakage for $G_3$ and $G_{80}$.&\\
\hline Galactose& Early 		& $G_3 \xrightarrow{F_3(G_i)} G_3^*$&Galactose activates $G_3$ molecules, which dimerize.&  \\
		& Induction&$G_3^*+G_3^* \xrightarrow{K_{D,3}} G_{3,d}^*$&Upon dimerization, these molecules bind to the full&\\
		&			&$D_2  +G_{3,d}^*\xrightarrow{K_{B,3}}
D_3$ &complex and induce transcription of Gal proteins.&\ref{fig:galregulon}(C)\\ \hline 
Galactose&Late 		&$G_1 \xrightarrow{F_1(G_i)} G_1^*$ 	& Activated
$G_1$ dimers replace $G_3$ dimers and&\\
		& Induction 	&$G_1^*+G_1^* \xrightarrow{K_{D,1}}
G_{1,d}^*$&
transcription continues. &  \ref{fig:galregulon}(D)\\
		& 			& $D_3 + G_{1,d}^* \xrightarrow{K_{B,1}}
D_4+G_{3,d}^*$ 				
		& &\\ \hline	
\end{tabular}
\caption[Schematic representation of the kinetic reactions occurring at the
level of the GAL mRNA promoter.]{Schematic representation of the kinetic reactions occurring at the
level of the GAL mRNA promoter. The promoter
conformations used in the ``Description" are $D_1=[UAS]_g:G_{4,d},$ 
$D_2=[UAS]_g:G_{4,d}:G_{80,d}]$,
$D_3=[UAS]_g:G_{4,d}:G_{80,d}:G_{3,d}^*$ and
$D_4=[UAS]_g:G_{4,d}:G_{80,d}:G_{1,d}^*$. $K_{D_i}$, $i\in \{80,3,1\}$ are the dissociation constants of the dimers into monomers, whereas $K_{B,i}$,
$i\in \{8,3,1\}$ are the dissociation constants of the promoter conformations. $F_3(G_i)$ and $F_1(G_i)$ are activation rates for $G_3$ and $G_1$ molecules in the presence of galactose.}
\label{table:table_rxn}
\end{center}
\end{sidewaystable}

Based on the above discussion, we conclude that the model should contain the following two reactions: the interactions of Gal3p ($G_3$) and Gal1p ($G_1$) with intracellular galactose ($G_i$) as determined by the two transitions
\begin{equation}
G_3 \xrightarrow{F_3(G_i)} G_3^*, \quad G_1 \xrightarrow{F_1(G_i)} G_1^*,
\label{eq:act}
\end{equation}
where $F_3$ and $F_1$ are the reaction rates appearing in Table \ref{table:table_rxn}, and $G_3^*$ and $G_1^*$ are the active forms of $G_3$ and $G_1$, respectively. The exact activation mechanism of these reactions has not been elucidated so far. Therefore, we assume here that they follow a saturating function with Michaelis-Menten kinetics:
\begin{equation}
\label{eq:act2}
F_k=\frac{\kappa_{C,k}G_i}{K_{S}+G_i},\quad k\in\{1,3\},
\end{equation}
where $\kappa_{C,1}$ and $\kappa_{C,3}$ are the maximal catalytic rates and $K_{S}$ is the galactose concentration for half-maximal activation.

There are various conformations in which the promoter can exist, based on the carbon source available to the cell. These conformations play a crucial role in determining the probability of gene expression. In our model, we will use ($\mathcal{R}$) to describe the probability of transcription, also known as the fractional transcription level, which is similar to that used in \cite{apostumackey} and \cite{venkatesh1999}. This quantity represents the fraction of the GAL promoters that is active and is expressed by
\begin{equation*}
\begin{array}{l}
\mathcal{R}_1=\frac{D_1+D_3+D_4}{D_1+D_2+D_3+D_4}=1-\frac{D_2}{D_1+D_2+D_3+D_4}=1-\frac{1}{1+\frac{K_{D,80}K_{B,80}}{G_{80}^2}+\frac{
(G_3^*)^2}{K_{D,3}K_{B,3}}+\frac{(G_1^*)^2}{K_{D,1}K_{B,1}K_{B,3}}},
\end{array}
\end{equation*}
whenever the promoter contains one single $[UAS]_g$. $D_1$, $D_2$, $D_3$ and $D_4$ are the four promoter conformations, as described in Table \ref{table:table_rxn} and Fig. \ref{fig:galregulon}, and $K_{D,i}$ and $K_{B,i}$ (with $i\in \{1,3,80\}$) are the dissociation constants obtained using quasi-steady state (QSS) assumptions on the dimerization and on the binding reactions. A complete derivation of $\mathcal{R}_1$ is given in \ref{app:ftl}. 

When the promoter, however, contains multiple $[UAS]_g$, the expression for the probability of transcription is significantly more complex:
\begin{equation}
\begin{array}{l}
{\mathcal{R} _{n} (G_{80} ,G_{3}^*,G_{1}^*)}{=1-\frac{1}{1+\sum\limits_{k=1}^n\Big(\frac{\sqrt{K_{D,80}K_{B,80}}}{G_{80}}\Big)^{2k}
+\sum\limits_{k=1}^{n}\Big(\frac{G_{3}^*}{\sqrt{K_{D,3}K_{B,3}}}\Big)^{2k} + \sum\limits
_{k=1}^{n}\Big(\frac{G_{1}^*}{\sqrt{K_{D,1}K_{B,1}K_{B,3}}}\Big)^{2k}},}
\end{array}
\label{eq:R}
\end{equation}
where $n$ is the total number of $[UAS]_g$, a quantity that is equivalent to the number of $G_4$ dimers binding at the GAL promoter site, as shown in Fig. \ref{fig:galregulon}.

Assuming that the activation reactions in Eqs. (\ref{eq:act}) are at QSS, we can write the fractional transcription level in Eq. (\ref{eq:R}) in terms of the non-activated proteins Gal3 and Gal1 as follows:
\begin{equation}\label{eq:Rss}
{\mathcal{R} _{n} (G_{80} ,G_{3} ,G_{1} ,G_{i} )}
{=1-\frac{1}{1+\sum\limits_{k=1}^n\Big(\frac{K_{80}}{G_{80}}\Big)^{2k}+\sum\limits
_{k=1}^{n}\left(\frac{G_{3} G_{i} }{K_3(K_{S}+G_{i})} \right)^{2k} + \sum\limits_{k=1}^{n}\left(\frac{G_{1} G_{i} }{K_1(K_{S} +G_{i})}\right)^{2k}}},
\end{equation}
where the constants $K_{80}$, $K_{3}$ and $K_{1}$ are given by
\begin{equation}\label{eq:Ks}
 K_{80}=\sqrt{K_{D,80}K_{B,80}}, K_{3}=\frac{\sqrt{K_{D,3}K_{B,3}}(\gamma_{G,3}+\mu_a)}{\kappa_{C,3}},K_{1}=\frac{\sqrt{K_{D,1}K_{B,3}K_{B,1}}(\gamma_{G,1}+\mu_a)}{\kappa_{C,1}}.
\end{equation}

From a dynamic point of view, galactose activates the synthesis of elements responsible for its own consumption. There are three important feedback loops in the gene network (regulated by Gal80p, Gal3p and Gal1p) and two others in the metabolic system (induced by Gal2p and Gal1p) that are part of galactose consumption. Considering their effect on GAL transcription, they constitute four positive feedback loops (involving Gal4p, Gal3p, Gal2p and Gal1p) and one negative feedback loop (involving Gal80p). 

As in the previous regulon model of \citet{apostumackey}, we do not include Gal4p in our modeling approach, since GAL4 transcription is neither repressed in glucose \citep{timson2007}, nor subject to the bistability property of the other Gal proteins \citep{acar2005}. This, as a result, leaves us with three important regulatory proteins of the gene network; namely, Gal3p, Gal80p and Gal1p.

As mentioned previously, Gal3p creates a positive feedback loop within the system. By letting $M_3$ denote the level of GAL3 mRNA, we can express its galactose-driven production in terms of $\mathcal{R}_1$ and write the dynamic rate of change for $M_3$ as
\begin{equation} \label{eq:m3}
\frac{dM_{3} }{dt} =\kappa _{tr,3} \mathcal{R} _{1} (G_{80} ,G_{3} ,G_{1} ,G_{i})-(\gamma _{M,3} +\mu_a)M_{3},
\end{equation}
where \textit{$\kappa_{tr,3}$} is the maximal transcription rate, \textit{$\gamma_{M,3}$} is the inherent cellular mRNA degradation rate and $\mu_a$ is the degradation rate due to dilution caused by cellular growth (i.e., the effective degradation rate for GAL3 is $\gamma_{M,3}+\mu_a$). We will use $\mu_a$ hereafter to represent the dilution rate of all molecular species. 

As described by \cite{abramczyk2012}, Gal3p is a ``ligand sensor"; upon activation, it binds to galactose molecules and subsequently removes the transcriptional inhibition exerted by Gal80p. Its mRNA promoter region is characterized by having a single binding site for the Gal4p dimer, having a leaky expression in raffinose and being fully expressed when galactose is present in the extracellular medium. Dynamically, GAL3 mRNA is translated at a rate $\kappa_{tl,3}$ and the associated protein  ($G_3$) is degraded at a rate $\gamma_{G,3}$ and its concentration is diluted at a rate $\mu_a$. A fraction of the Gal3p concentration is also activated by intracellular galactose ($G_i$), as described by Eq. (\ref{eq:act2}).  Thus,
\begin{equation}
\label{eq:g3}
\frac{dG_{3} }{dt} =\kappa _{tl,3} M_{3} -\left(\gamma _{G,3} +\mu_a +\frac{\kappa_{C,3} G_{i} }{K_{S} +G_{i} } \right)G_{3}.
\end{equation}

Gal80p is the inhibitory component of the GAL gene network. In the model,
$M_{80}$ represents GAL80 mRNA levels and $\kappa_{tr,80}$
and $\gamma_{M,80}$ denote its transcription and degradation rates, respectively. Based on this, we conclude that:
\begin{equation}\label{eq:m80}
\frac{dM_{80} }{dt} =\kappa _{tr,80} \mathcal{R} _{1} (G_{80} ,G_{3} ,G_{1} ,G_{i}
)-(\gamma _{M,80} +\mu_a )M_{80}.
\end{equation}

For the dynamic changes of the Gal80 protein, similar processes as those appearing in Eq. (\ref{eq:g3}) for the Gal3 protein are considered, except for the activation induced by galactose binding, which is absent here. Denoting GAL80 translation rate by $\kappa_{tl,80}$ and Gal80p degradation rate by $\gamma_{G,80}$, we obtain
\begin{equation}\label{eq:g80}
\frac{dG_{80} }{dt} =\kappa _{tl,80} M_{80} -(\gamma _{G,80} +\mu_a )G_{80}.
\end{equation}

The presence of four $[UAS]_g$ on the GAL1 promoter region implies that its transcription must depend on $\mathcal{R} _{4}$. By letting $\kappa_{tr,1}$ and $\gamma_{M,1}$ denote GAL1 transcription and degradation rates, respectively, the resulting equation governing $M_{1}$ dynamics is
\begin{equation}
\label{eq:m1}
\frac{dM_{1} }{dt} =\kappa _{tr,1} \mathcal{R} _{4} (G_{80} ,G_{3} ,G_{1} ,G_{i}
)-(\gamma _{M,1} +\mu_a )M_{1}.
\end{equation}

To describe the dynamic changes in Gal1p concentration, we will use  $\kappa_{tl,1}$  and $\gamma_{G,1}$ to denote GAL1 translation and Gal1p degradation rates, and use Michaelis-Menten kinetics of Eq. (\ref{eq:act2}) to describe its activation by $G_i$. Based on this, the rate of change of Gal1p is
\begin{equation} \label{eq:g1}
\frac{dG_{1} }{dt} =\kappa _{tl,1} M_{1} -\left(\gamma _{G,1} +\mu_a +\frac{\kappa
_{C,1} G_{i} }{K_{S} +G_{i} } \right)G_{1}.
\end{equation}

From a metabolic point of view, $G_1$ kinase converts ATP and galactose into ADP and Galactose-1-Phosphate ($G_p$). This phosphorylated form of galactose is known to inhibit the kinase, through a mixed inhibition reaction involving the $G_p$ product binding to the $G_1$ enzyme at an allosteric and an orthosteric positions on the enzyme \citep{timsonreece2002}. This process is included in the metabolic reactions described in the following section. 

\subsection{Metabolic network} \label{subsec:met}
\begin{table}
\centering
\begin{tabular}   {c l}  \hline
\vspace{-6pt}\\
\textbf{Processes}&\textbf{Description}\\ \vspace{-6pt}\\ \hline
$G_{e}\xrightarrow{T(G_2,G_i)} G_{i}$	&Galactose transport by $G_2$ across the cell membrane is a carrier-\\				&diffusion facilitated process \citep{ebel1985} and forms one of  \\
		&the positive feedback loops of the system.\\ \hline \vspace{6pt}
$G_{i} \xrightarrow{P(G_1,G_i,G_p)} G_p$	&$G_1$ phosphorylates $G_i$ and $G_p$ exerts a mixed inhibition on $G_1$.\\ \hline
\vspace{-6pt}
\end{tabular}
\caption{Simplified reactions of the metabolic branch. Galactose transport across the cell membrane is facilitated by $G_2$ permease. Intracellular galactose ($G_i$) is then metabolized through a phosphorylation reaction catalyzed by $G_1$ kinase.}
\label{tab:table_rxn2}
\end{table}

\begin{figure}[!htb]
\centering
\includegraphics[width=14cm]{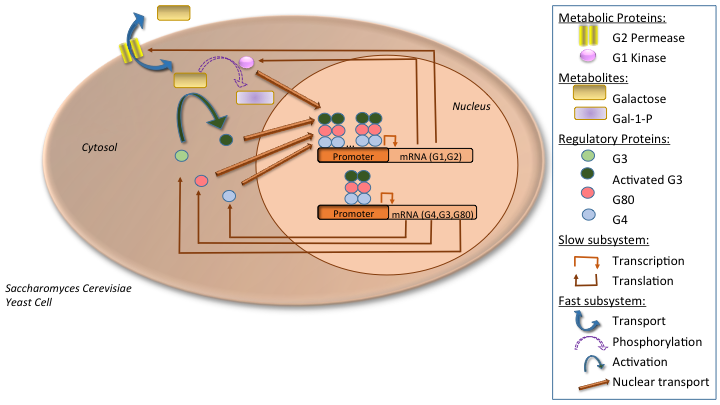}
\caption[Schematic illustration of the full galactose network]{Schematic illustration of the full galactose network, containing the genetic processes underlying galactose metabolism (to be read from the top-left corner). Via facilitated diffusion, galactose gets transported across the plasma membrane by the permease Gal2p ($G_2$). Intracellular galactose then activates the protein Gal3p ($G_3$), which dimerizes and binds to the complex $[G_{80}:G_{80}]:[G_4:G_4]$ found on the upstream activating sequence ($[UAS]_g$) on the promoter region of the GAL regulon. The newly formed complex allows RNA polymerase to start transcription of the GAL genes. These genes encode the regulatory proteins $G_3$, $G_4$, $G_{80}$, whose mRNA promoter regions contain a single $[UAS]_g$, as well as the metabolic proteins $G_2$ and $G_1$, which have, respectively, two and five $[UAS]_g$ \citep{sellickreece2008}. $G_3$ has a dual role, acting both as a kinase, phosphorylating galactose into galactose-1-phosphate (Gal-1-P) and also 
as a 
transcriptional activator \citep{abramczyk2012}. $G_1$, activated by galactose, dimerizes and replaces activated $G_3$ dimers, that are bound to $G_{80}$, allowing for further transcription to occur.}
\label{fig:scheme}
\end{figure}

\noindent 
In our model, we will include the following reaction steps of the Leloir pathway:
\begin{equation*}
G_e \stackrel{G_2(\alpha)}{\longleftrightarrow} G_i
\stackrel{G_1(\kappa _{GK} )}{\longrightarrow} G_p \stackrel{\text{Metabolism}(\delta
)}{\longrightarrow} \text{Glycolysis},
\end{equation*}
where $G_e$ and $G_i$ are the extracellular and intracellular galactose concentrations. The symbols above the arrows represent the reactions of the Leloir pathway, with the specific enzymes involved ($G_2$ and $G_1$) as well as the reactions rates (shown between parentheses) for galactose transport ($\alpha$), its phosphorylation rate ($\kappa _{GK}$ ) and $G_p$ consumption rate ($\delta$). These reactions are also shown in Fig. \ref{fig:scheme} and Table \ref{tab:table_rxn2}. Since $G_i$ is known to be a transcriptional activator of the Leloir enzymes downstream from $G_p$, it is reasonable to assume that, overall, $\delta$ represents these processes in the form of a negative feedback loop. As a result, $G_p$ is the last metabolite that we consider in our model.

Gal2p is a transmembrane, symmetric diffusion carrier. It is the main galactose transporter, followed by other hexose transporters, which form the HXT family. GAL2 mRNA has two $[UAS]_g$ for activation, implying that its probability of expression can be described by $\mathcal{R}_{2}$, with a maximum transcription rate $\kappa_{tr,2}$. The rates of change for the GAL2 and Gal2p species are
\begin{subequations}  \label{eq:mg2}
\begin{align}
\frac{dM_{2} }{dt} &=\kappa _{tr,2} \mathcal{R} _{2} (G_{80} ,G_{3} ,G_{1} ,G_{i}
)-(\gamma _{M,2} +\mu_a )M_{2}\\
\frac{dG_{2} }{dt} &=\kappa _{tl,2} M_{2} -(\gamma _{G,2} +\mu_a )G_{2},
\label{eq:m2g2}
\end{align}
\end{subequations}
where $\gamma_{M,2}$ and $\gamma_{G,2 }$ are the degradation rates for GAL2 mRNA and Gal2p, respectively, and $\kappa_{tl,2}$ is the translation rate.

As for the dynamics of the GAL1 mRNA and its associated protein, the last element considered in the metabolic network, they have already been discussed in the regulatory network, in Eqs. (\ref{eq:m1}) and (\ref{eq:g1}). 

As mentioned previously, galactose is transported via Gal2p by facilitated diffusion, with a maximal rate \textit{$\alpha$} and half-maximum transport rate $K$. The transport of the molecule ($T$) across the plasma membrane is completely governed by the balance between extracellular and intracellular concentrations, as follows
\begin{equation} \label{eq:transport}
T(G_2,G_i)=\alpha G_{2} \left(\frac{G_{e} }{K+G_{e} } -\frac{G_{i}
}{K+G_{i} } \right).
\end{equation}

This expression stems from a simplification of a process describing a carrier-facilitated diffusion
\citep{ebel1985} and is equivalent to the one found in \cite{atauri2005}.

Inside the cell, galactose is converted to a phosphorylated form (Gal-1-p, or $G_p$) via the Gal1p kinase. The phosphorylation is inhibited by the $G_p$ product through a mixed inhibition \citep{timsonreece2002, segal1970}, and is described by a Michaelis-Menten function, having a maximum rate $\sigma $ and a half-maximum activation $\kappa_{p}$, both dependent on the intracellular galactose concentration, as follows
\begin{subequations} \label{eqs:skp}
\begin{align}
\sigma (G_{i} )&=\frac{\kappa _{GK} K_{IU} K_{IC} }{K_{IU} K_{m} +K_{IC} G_{i} }
=\kappa _{GK} X \\
k_{p} (G_{i} )&=\frac{(K_{m} +G_{i} )K_{IU} K_{IC} }{K_{IU} K_{m} +K_{IC} G_{i}
} =(K_{m} +G_{i} )X, 
\end{align}
\end{subequations}
where $X=\frac{K_{IU}K_{IC}}{K_{IU}K_m+K_{IC}G_i}$. Galactose also activates Gal3 and Gal1 proteins in the cytoplasmic medium. Hence, by considering galactose transport, its phosphorylation, its activation of Gal1p and Gal3p, as well as its dilution ($\mu_a$), we can express the rate of change of this monosaccharide by the equation
\begin{equation} \label{eq:gi}
\frac{dG_{i} }{dt} =T(G_2,G_i)-\frac{\sigma (G_{i} )}{\kappa _{p} (G_{i} )+G_{p} } G_{1}
G_{i} -G_{i} \left(\frac{\kappa _{C,3} G_{i} }{K_{S} +G_{i} } +\frac{\kappa
_{C,1} G_{i} }{K_{S} +G_{i} } +\mu_a \right),
\end{equation}
where $T(G_2,G_i)$, $\sigma(G_i)$ and $k_p(G_i)$ are given by Eqs. (\ref{eq:transport}) and (\ref{eqs:skp}a-b), respectively.

The next metabolite in the Leloir pathway is $G_p$, the phosphorylated form of galactose. Since none of the compounds downstream of Gal1p in the metabolic pathway has a feedback on the regulatory processes, the remaining metabolic reactions have been approximated by a single consumption parameter $\delta$, as follows:
\begin{equation} \label{eq:gp}
\frac{dG_{p} }{dt} =\frac{\sigma (G_{i} )}{\kappa _{p} (G_{i} )+G_{p} } G_{1}
G_{i} - (\delta+\mu_a) G_{p}.
\end{equation}

\subsection{Galactose network model under glucose repression} \label{subsec:glu}
\noindent 
We intend to study how yeast cells adapt to environmental conditions where both galactose and glucose are available. Given that glucose is a repressor of the galactose network, we will examine how cells respond to an oscillatory glucose forcing in the presence of galactose. In order to do this, four repressive processes, based on experimental evidence, have been implemented.

\subsubsection{Cell growth}
\noindent
As mentioned previously, glucose is the energy source preferred by organisms, since they grow faster upon glucose rather than galactose exposure. According to the model, the dependence on the energy source is reflected in the dilution rate, which was previously denoted by $\mu_a$. We use an increasing Hill function to express the dependency of the growth rate, $\mu$, on glucose concentration, as follows:
\begin{equation} \label{eq:mu}
\mu (R)=\mu _{a} + \frac{\mu_{b}R^{n_\mu}}{\mu_c^{n_\mu}+R^{n_\mu}}, \quad n_\mu > 0,
\end{equation}
where $R$ is the glucose concentration, $\mu_a$ is the basal dilution rate, $\mu_a+\mu_b$ is the maximum dilution rate, $\mu_c$ is the half-maximum dilution and $n_{\mu}$ is the Hill coefficient.
\subsubsection{Transporter degradation}
\noindent
Experimental data indicates that glucose enhances vesicle degradation of the Gal2p transporter \citep{horakwolf,ramos1989}. To capture this effect, we assume here that the degradation rate of Gal2p ($\gamma_{G,2}$) follows a Hill function in its dependence on external glucose concentration according to the equation
\begin{equation} \label{eq:gamma}
\gamma _{G,2} (R)=\frac{\gamma _{b}R^{n_\gamma}}{\gamma_c^{n^\gamma} +R^{n_\gamma}}, \quad n_\gamma >0,
\end{equation}
where $\gamma_b$ is the maximum degradation rate, $\gamma_c$ is the half-maximum degradation and $n_{\gamma}$ is the Hill coefficient.
\subsubsection{Transcriptional regulation}
\noindent
Although the molecules involved in this process have been discovered, most of the research in this area focuses on presenting the overall reaction and the main factors without providing the necessary data for the quantification of the repression induced by glucose. Therefore, we approximate this process by 
\begin{equation} \label{eq:x}
x(R)=\frac{1}{\left(\frac{R}{x_{C} } \right)^{n_x} +1}, \quad n_x \geq 1,
\end{equation}
where $x_C$ is the half-maximum of this repressive process and $n_x$ is the Hill coefficient. It is important to point out that the inhibitor molecule Gal80p is not affected by this repression mechanism, as suggested by \citet{bhatbook}.
\subsubsection{Transporter competition}
\noindent
Gal2p is the main transporter of galactose and it is also
a high-affinity transporter for glucose \citep{Maier2002,reifenberg1997}. This means that both monosaccharides compete
for the same transporter. To incorporate this competition, we assume that the rate of galactose transport depends on
a scaling factor $y(R)$ which describes the probability of galactose transport across the transmembrane protein. This
scaling factor is assumed to decrease when glucose is present in the medium. Here, we choose a Hill function of degree 1 to describe $y(R)$, as follows:
\begin{equation} \label{eq:y}
y(R)=(1-y_b) +\frac{y_b}{y_c +R }, \quad n_y >0,
\end{equation}
where $(1-y_b)\in[0,1]$ is the basal probability of galactose transport in the absence of glucose and $y_c$ is the half-maximum transport repression by glucose.

\subsection{Complete mathematical model of the galactose network in the presence of glucose}
\subsubsection{Nine dimensional (9D) GAL model}
\noindent
In the galactose network, metabolic reactions occur on a faster time scale than the rates of change of the proteins \citep{reznik2013}. For example, transcription and translation occur with a time scale on the order of minutes, whereas transport via facilitated diffusion and phosphorylation occur at a rate greater than 500 times per minute \citep{atauri2005}. This implies that we can make a quasi-steady state (QSS) assumption on Eq. (\ref{eq:gp}) depicting the dynamics of the phosphorylated form of galactose ($G_p$). By solving for the  equilibrium concentration of this compound ($G_{p,(ss)}$) (shown in \ref{app:ftl}), we can then replace $G_p$ in Eq. {\ref{eq:gi}} by its steady state. This produces a nine dimensional model (9D) for the galactose network given by
\begin{subequations} \label{eqs:model-glu9}
\begin{align}
\frac{dM_{3} }{dt} &=\kappa _{tr,3}x(R) \mathcal{R} _{1} (G_{80} ,G_{3} ,G_{1} ,G_{i})-(\gamma _{M,3} +\mu(R)) M_3\\
\frac{dM_{80} }{dt} &=\kappa _{tr,80}x(R) \mathcal{R} _{1} (G_{80} ,G_{3} ,G_{1} ,G_{i})-(\gamma _{M,80} +\mu(R)) M_{80}\\
\frac{dM_{2} }{dt} &=\kappa _{tr,2}x(R) \mathcal{R} _{2} (G_{80} ,G_{3} ,G_{1} ,G_{i})-(\gamma _{M,2} +\mu(R)) M_2\\
\frac{dM_{1} }{dt} &=\kappa _{tr,1}x(R) \mathcal{R} _{4} (G_{80} ,G_{3} ,G_{1} ,G_{i})-(\gamma _{M,1} +\mu(R)) M_1\\
\frac{dG_{3} }{dt} &=\kappa _{tl,3}G_3-\left(\gamma _{G,3} +\mu
(R)+\frac{\kappa _{C,3} G_{i} }{K_{S} +G_{i} } \right)G_{3}\\
\frac{dG_{80} }{dt} &=\kappa _{tl,80}G_{80}-\left(\gamma _{G,80} +\mu
(R)\right)G_{80}\\
\frac{dG_{2} }{dt} &=\kappa _{tl,2} G_2-\left(\gamma _{2}(R) +\mu(R)\right)G_{2}\\
\frac{dG_{1} }{dt} &=\kappa _{tl,1} G_1- \left(\gamma _{G,1} +\mu
(R)+\frac{\kappa _{C,1} G_{i} }{K_{S} +G_{i} } \right)G_{1}\displaybreak[3]\\
\begin{split}
\frac{dG_{i} }{dt} &=\alpha y(R)G_{2} \left(\frac{G_{e} }{K+G_{e} }
-\frac{G_{i} }{K+G_{i} } \right)-\frac{2\sigma(G_i)G_iG_1}{k_p(G_i)+\sqrt{k_p(G_i)^2+\frac{4\sigma(G_i)G_iG_1}{\delta}}} -\\
&\quad -G_{i} \left(\frac{\kappa _{C,3} G_{i} }{K_{S} +G_{i} }
+\frac{\kappa _{C,1} G_{i} }{K_{S} +G_{i} } \right)-\mu (R)G_{i} ,
\end{split}
\end{align}
\end{subequations}
where
$\mathcal{R} _{n} (G_{80} ,G_{3} ,G_{1} ,G_{i} )$ is given in Eq. (\ref{eq:R}) and the functions $\sigma (G_{i} )$ and $k_{p} (G_{i} )$ in Eqs. (\ref{eqs:skp}a-b). Notice that, in the absence of glucose, we have, according to Eqs. (\ref{eq:mu})-(\ref{eq:y}), $x(R)=y(R)=1$, $\gamma(R)=\gamma_{G,2}$ and $\mu(R)=\mu_a$. 

\subsubsection{Five dimensional (5D) GAL model}
\noindent
The 9D model can be reduced to a five dimensional (5D) model by applying QSS approximation on the variables representing the various mRNA species of Eqs. (\ref{eqs:model-glu9}a-d), based on the fact that their degradation rates are one order of magnitude larger than those of their corresponding proteins. This 5D model is given by 
\begin{subequations} \label{eqs:model-glu}
\begin{align}
\frac{dG_{3} }{dt} &=\frac{\kappa _{tl,3} \kappa _{tr,3}x(R)}{\gamma _{M,3} +\mu(R)}\mathcal{R} _{1} (G_{80} ,G_{3} ,G_{1} ,G_{i})-\left(\gamma _{G,3} +\mu
(R)+\frac{\kappa _{C,3} G_{i} }{K_{S} +G_{i} } \right)G_{3}\\
\frac{dG_{80} }{dt} &=\frac{\kappa _{tl,80} \kappa _{tr,80}}{\gamma _{M,80} +\mu(R)} \mathcal{R} _{1} (G_{80} ,G_{3} ,G_{1} ,G_{i} )-\left(\gamma _{G,80} +\mu
(R)\right)G_{80}\\
\frac{dG_{2} }{dt} &=\frac{\kappa _{tl,2} \kappa _{tr,2}}{\gamma _{M,2} +\mu(R)}x(R)\mathcal{R} _{4} (G_{80} ,G_{3} ,G_{1} ,G_{i})-\left(\gamma _{2}(R) +\mu(R)\right)G_{2}\\
\frac{dG_{1} }{dt} &=\frac{\kappa _{tl,1} \kappa _{tr,1}}{\gamma _{M,1} +\mu(R)}x(R) - \left(\gamma _{G,1} +\mu
(R)+\frac{\kappa _{C,1} G_{i} }{K_{S} +G_{i} } \right)G_{1}\\
\begin{split}
\frac{dG_{i} }{dt} &=\alpha y(R)G_{2} \left(\frac{G_{e} }{K+G_{e} }
-\frac{G_{i} }{K+G_{i} } \right)-\frac{2\sigma(G_i)G_iG_1}{k_p(G_i)+\sqrt{k_p(G_i)^2+\frac{4\sigma(G_i)G_iG_1}{\delta}}} -\\
&\quad -G_{i} \left(\frac{\kappa _{C,3} G_{i} }{K_{S} +G_{i} }
+\frac{\kappa _{C,1} G_{i} }{K_{S} +G_{i} } \right)-\mu (R)G_{i}.
\end{split}
\end{align}
\end{subequations}

As before, the functions $\mathcal{R} _{n} (G_{80} ,G_{3} ,G_{1} ,G_{i} )$, $\sigma (G_{i} )$, $k_{p} (G_{i})$, $\mu(R)$, $\gamma_{2}(R)$, $x(R)$ and $y(R)$ are defined by Eqs. (\ref{eq:R}), (\ref{eqs:skp}a),  (\ref{eqs:skp}b), (\ref{eq:mu}), (\ref{eq:gamma}), (\ref{eq:x}) and (\ref{eq:y}), respectively. In the absence of repression, induced by glucose ($R$), the model will be called hereafter ``reduced 5D model", whereas in the presence of glucose-repression, it will be called the ``extended 5D model".

All model variations of the GAL network presented above have been implemented in {\tt XPPAUT} and {\tt MATLAB}, for further analysis and numerical simulations. Readers can refer to \ref{app:A} for more information on the software techniques employed.

\subsection{Model parameters}
\noindent
Parameter values of the models listed above have been mostly estimated using experimental data obtained from the same yeast strain and under similar laboratory conditions.

\begin{table} 
\small
\centering 
\begin{tabular}{lrll}
\hline
\vspace{-6pt}\\
\textbf{Symbol}&\textbf{Model}&\textbf{Definition [Units] }&\textbf{References}\\
&\textbf{Value}& 	& \\
 \vspace{-6pt} \\ \hline
$\kappa_{r3}$ 	&0.329 &$M_3$ transcription rate [$\frac{\text{copies}}{\text{cell}\times\text{min}}$]& Calculated\\
$\kappa_{r80}$ 	&0.147 &$M_{80}$ transcription rate [$\frac{\text{copies}}{\text{cell}\times\text{min}}$]&\\
$\kappa_{r2}$ 	&0.678 &$M_2$ transcription rate [$\frac{\text{copies}}{\text{cell}\times\text{min}}$]&   \\
$\kappa_{r1}$ 	&1.042 &$M_1$ transcription rate [$\frac{\text{copies}}{\text{cell}\times\text{min}}$]&  \\ \hline
$\kappa_{l3}$	&645 &$G_3$ translation rate [$\frac{\text{molecules}}{\text{copies}\times\text{min}}$]& Calculated \\
$\kappa_{l80}$	&210 &$G_{80}$ translation rate [$\frac{\text{molecules}}{\text{copies}\times\text{min}}$]& \\
$\kappa_{l2}$ 	&800& $G_2$ translation rate [$\frac{\text{molecules}}{\text{copies}\times\text{min}}$] &\\
$\kappa_{l1}$	&187&$G_1$ translation rate [$\frac{\text{molecules}}{\text{copies}\times\text{min}}$] & \\  \hline
 c		&4.215$\times 10^7$&Conversion constant [$\frac{\text{molecules}}{\text{cell}\times\text{mM}}$]& Calculated\\ \hline
$\mu_a$ 	&4.438$\times 10^{-3}$&Dilution rate [min$^{-1}$] & \cite{tyson1979}\\
$\gamma_{M}$ 	&4.332$\times10^{-2}$ &$M_3$ degradation rate [min$^{-1}$]&  \cite{holstege1998},\\
& &  &\cite{bennett2008}\\
$\gamma_{G,3}$  &7.112$\times10^{-3}$&$G_3$ degradation rate [min$^{-1}$]& \cite{ramsey2006}\\
$\gamma_{G,80}$ &2.493$\times10^{-3}$&$G_{80}$ degradation rate  [min$^{-1}$]& \\
$\gamma_{G,2}$  &0&$G_2$ degradation rate  [min$^{-1}$] &  \\
$\gamma_{G,1}$  &0&$G_1$ degradation rate  [min$^{-1}$] & \\ \hline
$K_{D,80}$ 	&3$\times10^{-7}$& $G_{80_d}$ dissociation constant [mM] & \cite{melcher2001}\\
$K_{B,80}$ 	&5$\times10^{-6}$& $D_2$ dissociation constant [mM] &\cite{lohr1995} \\
$K_{B,3}$ 	&6$\times10^{-8}$& $D_3$ dissociation constant [mM]& \cite{acar2005}\\ \hline
$K_{B,1}$ 	&6$\times10^{-8}$&$D_4$ dissociation constant [mM]& Model Estimation\\
$K_{D,3}$ 	&$1.25\times10^{-2}$& $G_{3_d}^*$ dissociation constant [mM] & \\
$K_{D,1}$ 	&1& $G_{1_d}^*$ dissociation constant [mM]& \\
$K_{S}$ & $4000$ & $G_3$ and $G_1$ half-maximum activation& \\
& &  constants [mM]& \\ \hline
$\kappa_{C,3}$ &0.5&$G_3$ activation rate [$\frac{1}{\text{mM}\times\text{min}}$]& Model Estimation\\
$\kappa_{C,1}$ &$8\times10^{-5}$&$G_1$ activation rate [$\frac{1}{\text{mM}\times\text{min}}$]&\\ \hline
$\alpha$&4350& Maximum rate of symmetric&\cite{atauri2005} \\
& & facilitated diffusion [min$^{-1}$]& \\
$\kappa_{GK}$&702 &Experimentally measured phosphory-&\cite{brink2009}\\
& &lation rate of $G_i$ [min$^{-1}$] &\\
$\delta$&59200&Rate of Gal1p metabolism [min$^{-1}$]& \cite{atauri2005}\\ \hline		
$K$&1& Half-maximum concentration &\cite{atauri2005} \\
&	&for the transport process [mM]& \\
$K_m$& 1.2&Half-maximum concentration & \cite{timsonreece2002}\\
  & 		&for phosphorylation [mM]&\\
$K_{IC}$&160&Competitive inhibition  constant [mM]&\cite{timsonreece2002} \\
$K_{IU}$&19.1&Uncompetitive inhibition constant [mM]& \cite{timsonreece2002}\\ \hline
 \end{tabular}
\caption[Values of the model parameters of the galactose network]{Values of the model parameters of the galactose network. References are provided when the exact values of these parameters have
been measured or calculated from experimental data.} 
\label{table:par-estim} 
\end{table}

\subsubsection{Galactose parameters}
\noindent
The rates contained in Table \ref{table:par-estim} are: (a)~transcription and translation rates for all the four GAL mRNAs; (b)~degradation and dilution rates for all intracellular species; (c)~dissociation constants of compounds involved in the genetic regulation and (d)~metabolic rates including transport and phosphorylation. Whenever possible, these values are chosen to fit experimental and/or literature data. Calculations and detailed derivations of these results can be found in the subsection that follows. As for the constants involved in galactose-induced activation, they are estimated using numerical simulations under the assumption that the dynamics of the model must exhibit bistability. Parameter values listed in Table \ref{table:par-estim} are representative of a wild-type cell.

\subsubsection{Glucose parameters}

\begin{table} 
\centering 
\begin{tabular}{lrcl}
\hline
\vspace{-6pt}\\
\textbf{Symbol}&\textbf{Value}&\textbf{Units}&\textbf{Definition}\\
 \vspace{-6pt}\\ \hline
$\mu_b$ &0.00512 & min$^{-1}$ &Dilution rate in glucose\\
$\mu_c$ &0.3611 & \% w/v&Half-maximum constant for dilution \\
$n_\mu$ &1 & unitless&Hill coefficient for dilution \\ \hline
 $\gamma_b$&0.001416 & min$^{-1}$ &Gal2p degradation rate  \\
$\gamma_c$ &0.8592 & \% w/v &Half-maximum constant for degradation\\
$n_\gamma$ &1 & unitless&Hill coefficient for Gal2p degradation \\    \hline
$x_c$ &0.2443&\% w/v &Half-maximum constant for transcriptional regulation \\
$n_x$ &1 & unitless&Hill coefficient for transcriptional regulation\\ \hline
$y_b$ &0.0003& min$^{-1}$ &Increase in the competition rate due to repression\\
$y_c$ &2.9989 & \% w/v &Half-maximum activation for repressive competition \\
$n_y$ &1 & unitless &Hill coefficient for the competition between\\
& & &glucose and galactose for the Gal2p transporter\\    \hline
 \end{tabular}
\caption[Kinetic parameters of glucose repression]{Kinetic parameters of glucose repression in the extended 5D model. ``Cftool" was used to fit the functions describing
dilution and $G_2$ transporter degradation, whereas the genetic algorithm was used to fit the parameters involved in the last two processes of glucose repression.}
\label{table:par-glu} 
\end{table}

\noindent
In the next section, we present the results obtained from investigating the 5D and the 9D models that include the crucial repressive processes induced by glucose uptake. The parameters used in the modeling are obtained by fitting the mathematical expression of these repressive processes to experimental data using the ``Cftool" toolbox and the Genetic Algorithm (see Table \ref{table:par-glu}). The Hill coefficients $n_\mu$ and $n_\gamma$ are set to 1, to provide an ideal Michaelis-Menten relation representing the effect of glucose on cellular growth and degradation of the Gal2p transporter. The other two glucose-induced repressive processes (i.e., transcriptional repression and transporter competition), require more complex fitting procedures. As done before, the coefficients $n_x$ and $n_y$ are chosen to be either 1 or 2. In a qualitative sense, these coefficients represent the sensitivity of the process to glucose concentration. The minimizing error of the Genetic Algorithm gave a better result for the value of the coefficients being 1, as shown in Table \ref{table:par-glu}.
\section{Results} 
\label{results}
\noindent
Given the complexity of the 9D model, we first focus here on the dynamics of the reduced and extended 5D models. We begin by examining how bifurcation structure of both these models is altered in response to changes in biological quantities that can be manipulated in a physical setting. By doing so, we can draw close connections between predicted model behaviours and observed experimental results. We will then examine the temporal response of the extended 5D model to periodic forcing by extracellular glucose to elucidate the low-pass filtering nature of the GAL network as suggested by \citet{bennett2008}.

\subsection{Bistability with respect to galactose}
\subsubsection{Steady state behaviour of metabolic proteins}
\noindent
To examine how the model depends on extracellular galactose and glucose, we study how the steady state behaviour of the (reduced and extended) 5D models is modified upon changes in the extracellular concentration of these two monosaccharides. The 5D model described by Eqs. (\ref{eqs:model-glu}a-e) is used for this purpose, since the QSS approximation assumed on the mRNA level in this model will not alter its steady state properties.

To conduct this analysis, a physiological range for extracellular  galactose $G_e$ is specified. This can be done by using galactose induction curves obtained under various experimental conditions (such as the type of strain and growth conditions used). For example, \citet{acar2010} showed that the diploid strain MA0496 exhibits GAL1 induction with 0.05 \% w/v galactose, whereas \citet{bennett2008} found that the haploid strain K699 is more sensitive and requires only 0.002 \% w/v for the same increase in induction levels. It was also demonstrated that the full activation of the GAL network occurs at 0.025 \% w/v for the wild-type strains used in \citet{acar2010} and \citet{venturelli2012}, and at 0.05\% w/v for the ones cited in \citet{bennett2008}. Given the extensive data available, we focus our analysis only on the wild type strain K699 and as a result choose the range of 0-0.2\% w/v for $G_e$. This is achieved by first plotting the one-parameter bifurcation of various metabolites in the reduced 5D model with respect to $G_e$ within this range.

\begin{figure}[!htb]
\hspace{-1cm}
\includegraphics[width=16cm]{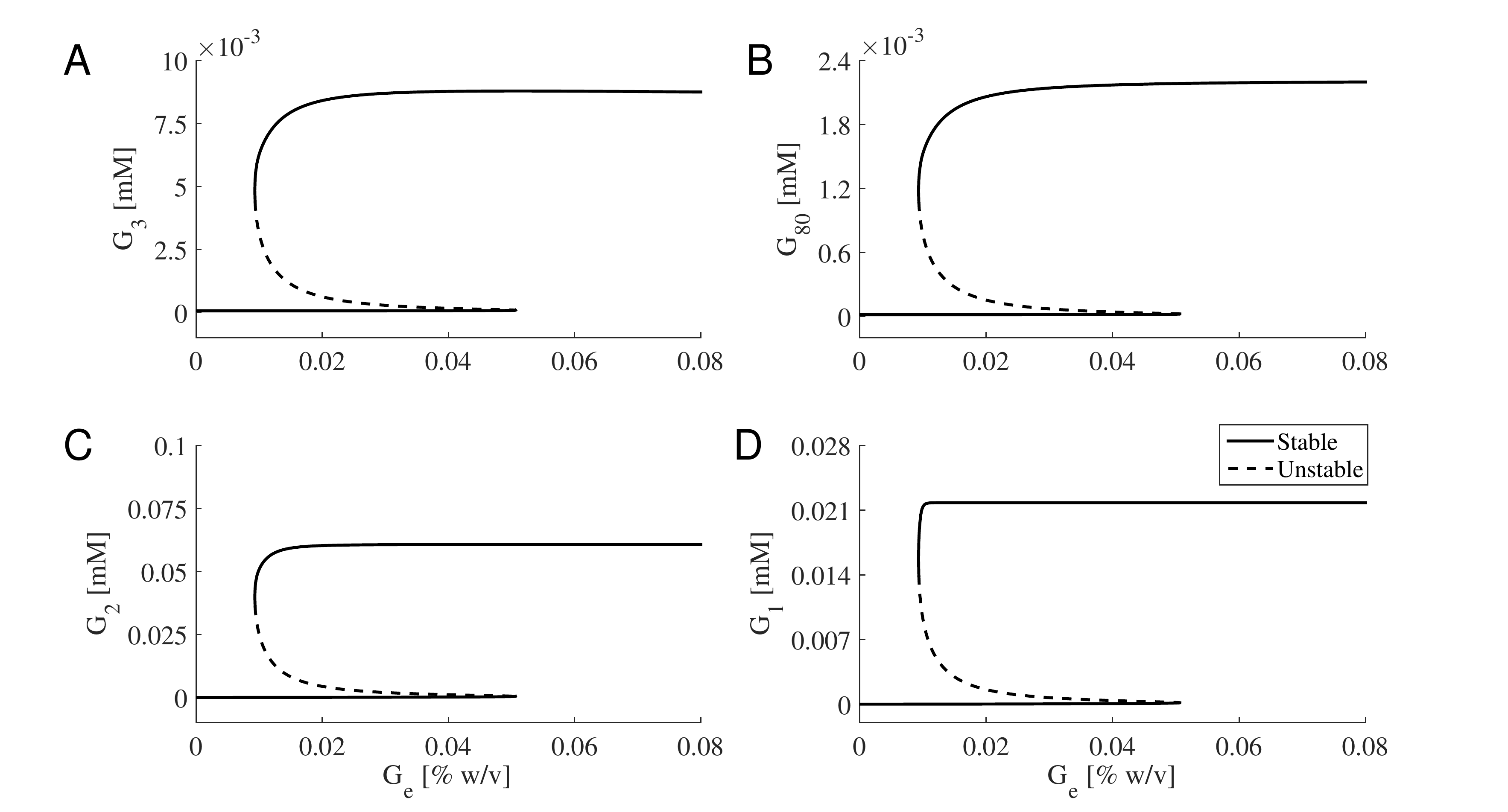}
\caption{One-parameter bifurcation of various proteins as a function of the extracellular galactose concentration ($G_e$), measured in units of weight/volume ([\% w/v]). The four panels show the steady state values of (A) the regulatory protein Gal3 ($G_3$); (B) the inhibitory regulatory protein Gal80 ($G_{80}$); (C) the permease Gal2 ($G_2$); and (D) the regulatory and enzymatic protein Gal1 ($G_1$). Black thick lines refer to the stable branches of attracting equilibria, whereas dashed lines represent the unstable branches of equilibria, separating the two stable branches within the bistable regime.}
\label{fig:result1}
\end{figure}

Figure \ref{fig:result1} shows model outcomes of the equilibrium concentrations associated with the four main Gal proteins: $G_3$ (panel A), $G_{80}$ (panel B), $G_2$ (panel C) and $G_1$ (panel D), as a function of $G_e$. In all cases, bistability is exhibited by the four variables in the form of two branches of attracting equilibria (black lines) that overlap over a wide range of values for $G_e$, separated by an unstable branch of equilibria (i.e., a branch of physiologically unattainable steady states). The right saddle node at the intersection of the unstable and the stable branches of Fig. \ref{fig:result1} has a numerical value of $0.05$ \% w/v. It represents the concentration of $G_e$ that produces full induction of the network, consistent to that of K699 yeast strain determined by \citet{bennett2008}.

The ratio between the uninduced and the induced states occurring in the bistable regime is comparable to those ratios observed experimentally in \citet{acar2005}, \citet{acar2010} and \citet{venturelli2012} for the Gal1 and Gal10 mRNA promoter tagged with Yellow Fluorescence Protein (YFP). According to the reduced 5D model, the regulatory proteins Gal3 and Gal80 exhibit a ratio of 65 to 100 between the induced and uninduced states of their protein level, unlike the metabolic proteins Gal2 and Gal1, which exhibit a high ratio of 144 to a maximum of either 4320 or 6852 for the two types of metabolic proteins, respectively. The range of these ratios is governed indirectly by the number of upstream activating sequence at the promoter levels, which dictates the index of the fractional transcriptional level defined by Eq. \ref{eq:Rss}. Experimentally, the results for these ratios vary between laboratories: \citet{acar2005} showed an average of 33 to 37.5 fold difference between induced and uninduced YFP-tagged GAL1 promoter in wild type cells with a bistable regime between $0.06-0.3$ \% w/v of galactose. In another paper by the same group, the range of bistability was shown to be between $0.1-0.4$ \% w/v of galactose and the ratio of induced to non-induced states to range from 40 up to 330 for the same promoter type \citep{acar2010}. On the other hand, \citet{venturelli2012} recently measured the occurrence of the Gal10 promoter by YFP and observed a ratio of 30 to 100. The apparent discrepancy in the upper bound of this ratio between those obtained experimentally and our numerical results (see \ref{app:A}) maybe due to the fact that transcriptional and translational rates were estimated using experimental data obtained from different yeast strains \citep{arava2003,lashkari1997,ideker2001,bonven1979}. It could 
be also due to the stochastic nature of the data acquired using cultures that contained many cells, unlike our numerical results that are generated using deterministic single-cell models.

Overall, these results reveal that the bistability property studied in \citet{apostumackey} is not only conserved in our GAL network, but also it is an inherent property of the GAL regulon rather than the metabolic subnetwork. They also indicate that the induced-to-uninduced ratios are in agreement with certain experimental studies \citep{acar2005,acar2010,venturelli2012} and that this ratio is most sensitive to perturbations in the transcriptional and translational rates. 

\subsubsection{Steady state behaviour of intracellular galactose}

\begin{figure}[!htb]
\hspace{-1cm}
\includegraphics[width=18cm]{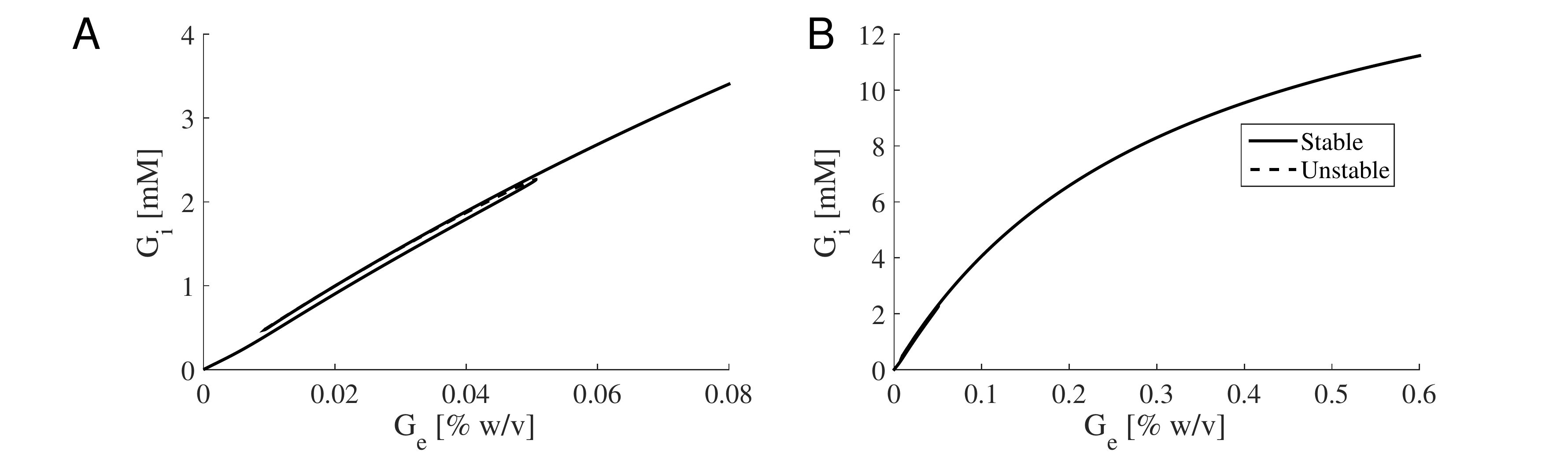}
\caption{One-parameter bifurcation with respect to the extracellular galactose concentration ($G_e$). The stable (solid) and unstable (dashed) branches of steady-state values of $G_i$ within the range (A) [0,0.08] \% w/v,  and (B) [0,0.6] \% w/v of galactose ($G_e$) as defined in Fig. \ref{fig:result1}. Due to the difference in time scales between the fast metabolic subsystem and the slow GAL regulon, the bistability in intracellular galactose $G_i$ is not as pronounced as that seen for the proteins, shown in Fig. \ref{fig:result1}.}
 \label{fig:result2}
\end{figure}
\noindent
As was done in the previous section, we extend the bifurcation analysis conducted in the proteins to intracellular galactose ($G_i$). The goal is to determine how the steady state behaviour of $G_i$ in the reduced 5D model depends on $G_e$ and if bistability is maintained. Fig. \ref{fig:result2}(A) shows that the bifurcation diagram of $G_i$ with respect to $G_e$, using the same range of [0, 0.08] w/v of galactose as that used in Fig. \ref{fig:result1}, also exhibits bistability in the form of a switch, but it is not as pronounced as those in Fig. \ref{fig:result1}. Indeed, the two stable branches of this bifurcation diagram are so close, they appear as one curve possessing a Hill-like profile that eventually plateaus at high values of $G_e$. The rate of change of $G_i$, according to the reduced model, depends on reactions that possess both slow and fast time scales. Due to the fast reactions taking place in the metabolic system (when compared to those in the GAL regulon), such as 
transport and phosphorylation, the bistable switch associated with $G_i$ is not as pronounced as that for the proteins of Fig. \ref{fig:result1}. Thus verifying this switching behaviour at the level of intracellular galactose could be experimentally challenging.

Nonetheless, it is important to point out that the switching behaviour in Fig. \ref{fig:result2} can occur for various reasons. The prominent change in stability occurring in the GAL network destabilizes the positive and the negative feedback elements of the system. Moreover, a change in the ratio between the transporter Gal2 and the enzyme Gal1, or a change in the concentration of the two activators, Gal1 and Gal3, could underlie this behaviour. As a result, we believe that bistability in the metabolic pathways of the GAL network occurs due to the inherent properties of the system at the gene regulation level. We also expect, based on the discussion above, the equilibrium concentration of $G_i$ to reach a saturating level at high extracellular galactose levels, predicted to occur around 0.5 \% w/v of $G_e$.

\subsubsection{Two-parameter bifurcations as a measure of sensitivity}
\noindent
To assess the sensitivity of the bistable regime changes (or perturbations) in the parameter values that are representative of variations in yeast strains, we study here how the two saddle nodes of Figs. \ref{fig:result1} and \ref{fig:result2} are affected by changes in the other rates of the system and how they alter the range of the bistable regime. These changes could reflect yeast strain variability due to genetic mutations which can create different functional properties or different growth rates and can either hinder or induce reactions by varying external factors. Plotting two-parameter bifurcations, that trace the location of these saddle nodes in a two dimensional parameter space, can help illustrate these variations and their effects.

\begin{figure}[!htb]
\hspace{-1cm}
\includegraphics[width=18cm]{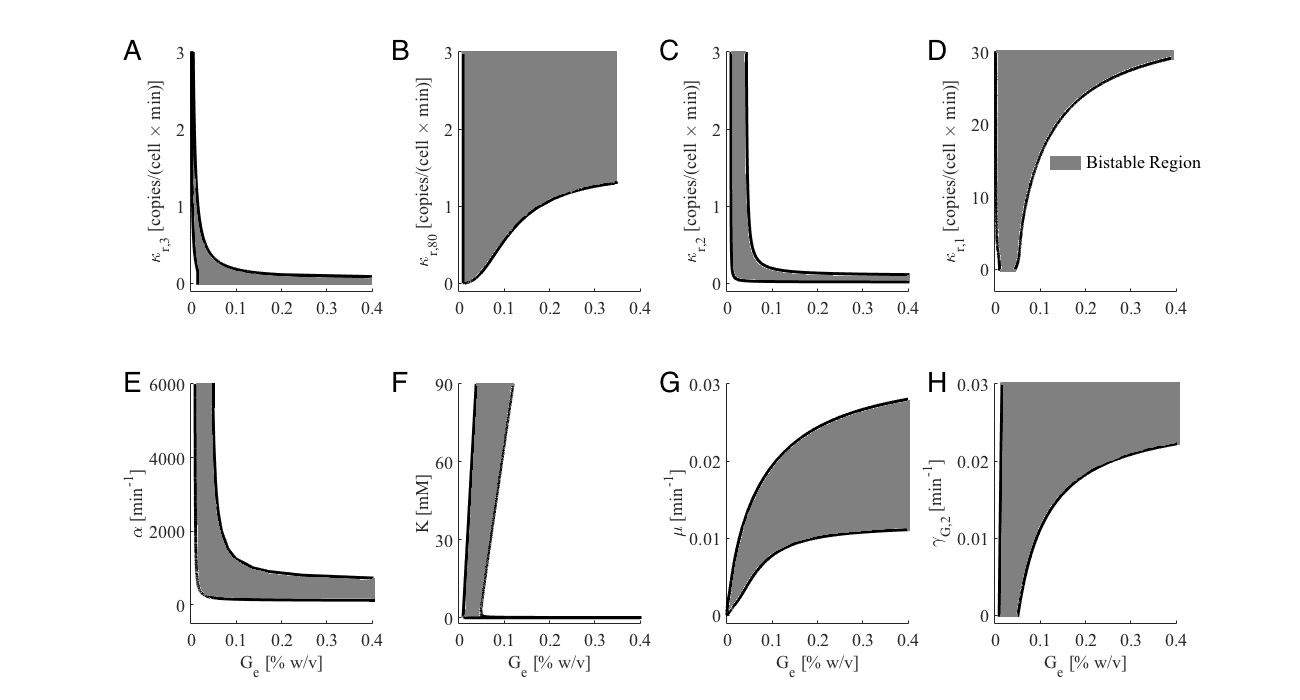}
 \caption{Two-parameter bifurcations of the reduced 5D model with respect to extracellular galactose ($G_e$) and other kinetics parameters of the model. These include (A) Gal3 transcription rate ($\kappa_{r,3}$); (B) Gal80 transcription rate ($\kappa_{r,80}$); (C) Gal2 transcription rate ($\kappa_{r,2}$);  (D) Gal1 transcription rate ($\kappa_{r,1}$); (E) Gal2-dependent galactose transport rate ($\alpha$); (F) half-maximum transport constant ($K$); (G) dilution rate ($\mu$), due to cellular growth; and (H) Gal2 degradation rate ($\gamma_{G,2}$). Each panel depicts the limit points (black lines), along with the bistable (grey) and the monostable (white) regimes. Notice the presence of the cusp in panels A and B, the dependence of the left limit point on $G_e$ in panel F, and the disappearance of the right limit point in panels B, D, G and H. }
\label{fig:result3}
\end{figure}

We begin first by considering the two-parameter bifurcations that uncover how transcriptional repression of the Gal proteins affects the bistability regime. Figure \ref{fig:result3}(A-D) displays in grey (white) the regimes of bistability (monostability) bordered by black lines that determine the location of the left and right limit points (or saddle nodes) of Fig. \ref{fig:result1}. The monostable (white) regimes could either correspond to the induced state (to the right of the grey regimes) or uninduced state (to the left of the grey regimes). As shown, a decrease in the transcriptional rates of Gal3 and Gal2 proteins, involved in positive feedbacks, can extend this regime (panels A and C, respectively) by shifting the right limit point further to the right. A major decrease in the transcription rate of Gal2, however, can eventually shift the system into the monostable regime that possesses the uninduced state, provided that $G_e$ is small enough. The transcription rates of Gal80 and Gal1 (both of which are 
self-inhibitory 
proteins), on the other hand, act in an opposite fashion (panels B and D, respectively). These results suggest that different mutants can have different dynamic properties. This may explain why bistability was not observed by all research groups. Since yeast cultures are heterogenous, each cell culture can be described by a particular parameter set of the mathematical model. Therefore, different cultures may belong to different stability regimes of the two-parameter bifurcation. 

We similarly proceed to investigate the dynamics of the reduced 5D model in response to other parameter variations of the model, representing potential mutations, in Fig. \ref{fig:result3}(E-H). More specifically, in Fig. \ref{fig:result3}(E), we investigate the effect of GAL2 genetic mutation that affects the functionality of this transporter by varying its galactose transport rate, $\alpha$. Our results reveal that when the rate of transport is higher than that of wild type with its default value listed in Table \ref{table:par-estim}, little effect on bistability is observed. However, once the transport is impeded, bistability regime broadens and the fully induced state may become unattainable (depending on initial conditions).

The bistability regime of the GAL network is also dependent on how rapidly yeast cells grow. Indeed, Fig. \ref{fig:result3}(G) shows that a lower growth rate (i.e. higher dilution rate $\mu$ than its default value) leads to a wider bistable regime. Similar results are observed in Fig. \ref{fig:result3}(H) when increasing the degradation rate of the transporter, a process that is regulated by ubiquitin ligase complexes \citep{horak2005}: in this case, we see an increase in the bistability regime of the GAL system subsequent to a decrease in the capacity of the cells to be fully induced, equivalent to an increased effect of the inhibitory proteins.

In all of these cases discussed above, the two limit points of the two-parameter bifurcations (i.e., the boundaries of the bistable grey regimes) are present, with the left one mostly remaining stationary at one specific concentration of extracellular galactose $G_e$. The two-parameter bifurcation associated with the half-maximum transport constant $K$ is the only one that does not follow the same pattern. Indeed, Fig. \ref{fig:result3}(F) shows that increasing $K$ causes the left limit point to shift to the right, increasing the width of the monostable regime associated with the uninduced steady state. This could be beneficial for the cell as it may allow it to compensate for problems in the galactose induction of certain yeast strains, a potentially recurrent phenomena in other eukaryotic cells.

\begin{figure}[!htb]
\centering
\includegraphics[width=9cm]{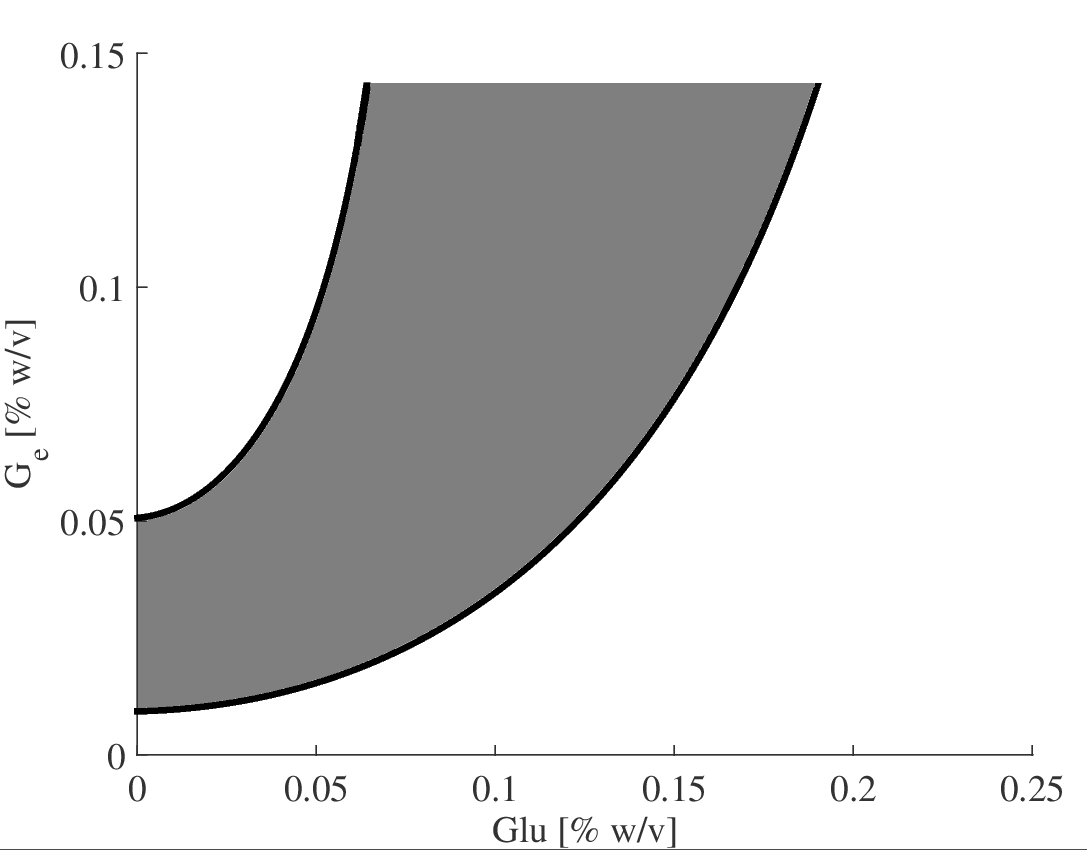}
\caption[Two parameter bifurcation with respect to extracellular galactose ($G_e$) and glucose ($R$).]{Two parameter bifurcation with respect to extracellular galactose ($G_e$) and glucose ($R$), for a yeast strain which shows the bistable regime (in grey) bounded by the limit points as defined in Fig. (\ref{fig:result3}).}
\label{fig:galglubif}
\end{figure}

It is important to point out that the bistability regime with respect to the inducer (i.e., to extracellular galactose), produced by the extended 5D model, is also sensitive to the repressor (i.e. glucose). Figure \ref{fig:galglubif} portrays this sensitivity as a two-parameter bifurcation, with respect to $G_e$ and glucose ($R$), which is qualitatively similar to the one seen in \cite{venturelli2015} and to the landscape diagram of \cite{stockwell2015}. With the parameter combinations shown in Tables \ref{table:par-estim} and \ref{table:par-glu}, we predict that the bistability property will persist even when no repressor ($R$) is present, a feature not mentioned in \cite{venturelli2015}. In an experimental setting, we expect the system to show a bistable response (also commonly called binary response) if the administered glucose concentration is $\sim 0.075$ \% w/v and $G_e$ is higher than 1 \% w/v. 

\subsection{Temporal response to an oscillatory glucose input}
\noindent
One important aspect of the GAL network established experimentally is its low-pass filtering capacity when wild-type and GAL2 mutant yeast cells, grown on a background medium of 0.2 \% w/v galactose, are subjected to a periodic glucose forcing of amplitude 0.125 \% w/v and a baseline of 0.125 \% w/v \citep{bennett2008}, identical to that shown in Fig. \ref{fig:result6}(A). The two main features associated with this filtering capacity is the decline in the amplitude and phase of the output response (both determined experimentally by measuring GAL1 mRNA expression level). Here, we analyze this phenomenon using the extended 5D and 9D GAL models, to determine whether this behaviour is captured by both.

\subsubsection{Dynamics of wild-type cells}
\begin{figure}[!htb]
\hspace{-1cm}
\includegraphics[width=19cm]{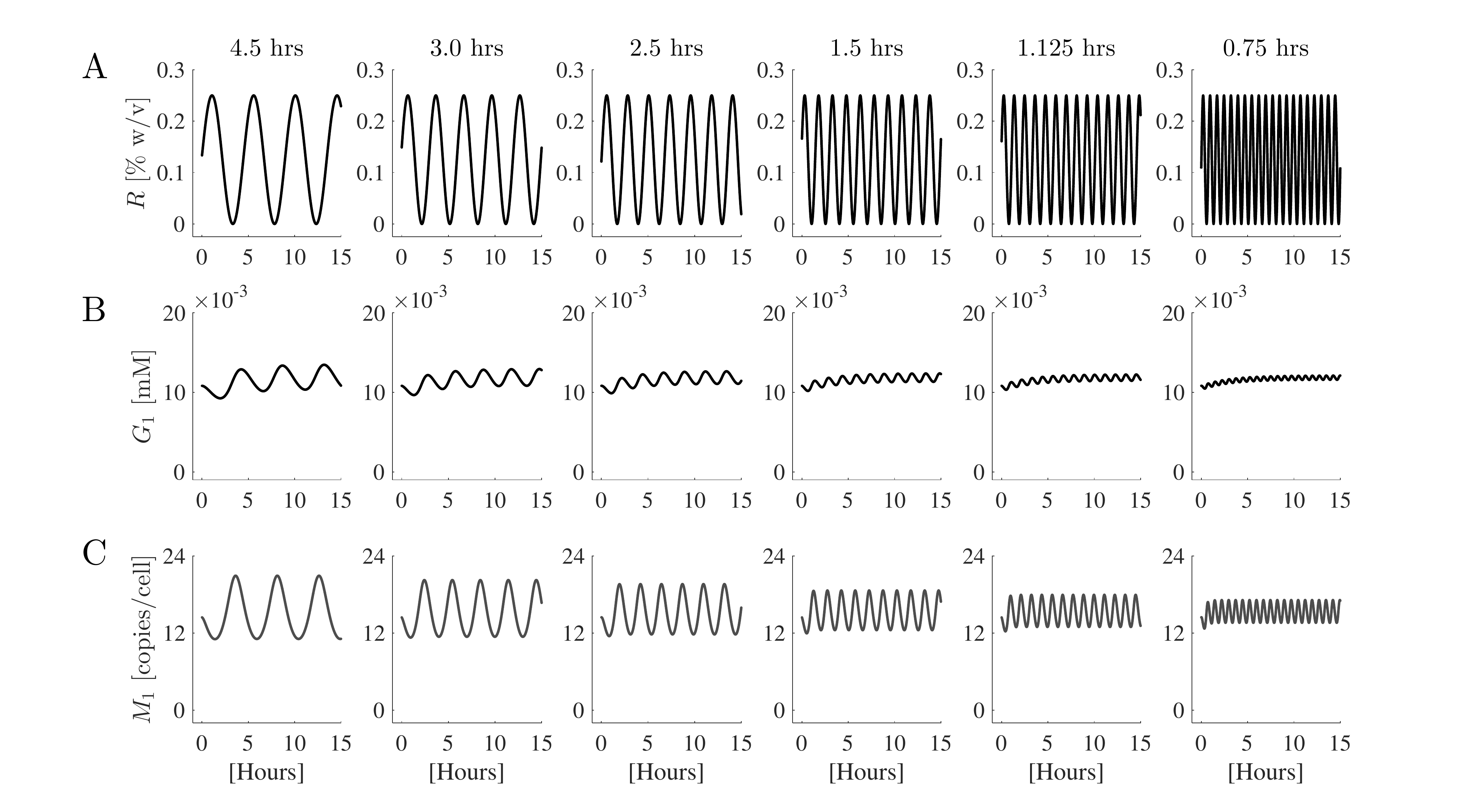}
\caption{The extended 5D model response to oscillatory glucose input signal of varying periods. (A) Extracellular glucose forcing with a period of 4.5, 3.0, 2.25, 1.5, 1.125 and 0.75 hours is applied on the GAL network. (B) Due to the changes in the external medium, the Gal1 output signal, as determined by the extended 5D model, shows that the system adapts after a transient period of 5 hours. (C) Using GAL1 mRNA ($M_1$) as an output signal of the 9D model, the adaptation time is only an hour, occurring this time upstream of the protein, at the transcriptional level.}
\label{fig:result6}
\end{figure}

\noindent
Using the extended 5D model, the output of the GAL network of the wild-type strain (corresponding to the default parameter values of the model) in response to glucose oscillations is shown in Fig. \ref{fig:result6}(B). As suggested above $G_1$ concentration is used in this figure to assess the oscillatory response of the system in terms of both amplitude and frequency. As shown, the amplitude of the $G_1$ output signal decreases when the frequency of the glucose input signal increases (i.e., they are inversely correlated with each other). This behaviour is accompanied by a transient period of 5 hours in which the output signal gradually ascents to an elevated baseline while oscillating. Our numerical results reveal that multiple factors associated with model dynamics and structure are generating such outcomes. In particular, bistability and the phase of the glucose input signal are the two key factors causing the elevation in the baseline, whereas  the presence of various time scales within the model is responsible for making the inverse correlation between the amplitude of output signal and the frequency of the input signal. It should be noted here that, similar to our numerical results, experimental recordings of the GAL1 mRNA output signal (which is upstream from $G_1$) shown in \citet{bennett2008} also displays similar ascent in the baseline, but with a shorter transient of around 1 hour. This suggests that although the reduced model possesses the core structure of the GAL network, the applied QSS assumptions seemed to undermine the ability of the model to capture the proper length of the transient.

Due to the presence of discrepancy in the transients between experimental and numerical results (obtained from the extended 5D model), we turn our attention now to the 9D model to analyze the effects of QSS assumption on its response to oscillatory glucose input signal. We do so by plotting the GAL1 mRNA expression level as an output signal of the model when the oscillatory glucose input signal of Fig. \ref{fig:result6}(A) is applied. Figure \ref{fig:result6}(C) shows that the inverse correlation between amplitude and frequency is preserved by the 9D model, and that the oscillations in the output signal exhibit higher peaks and more pronounced mRNA production during glucose decline in each cycle of the input signal. The figure also shows that the transient occurring before the oscillations in GAL1 mRNA reach a baseline lasts about 1 hour, which is consistent with the value observed experimentally \citep{bennett2008}. These results indicate that the 9D model is necessary when analyzing the temporal and transient dynamics of the system.

\subsubsection{Dynamics of mutant cells}
\noindent
As a test to validate the model against experimental data, we examine the predicted response of a GAL2 mutant strain (GAL2$\Delta$) to periodic external glucose forcing. \citet{bennett2008} described such mutant as requiring ten times more galactose for full induction than the wild-type strain. To capture this effect in our simulations, the mutant is modeled by varying the functional properties of $G_2$, i.e., by decreasing its transport rate $\alpha$ from 4350 to 702 min$^{-1}$. As demonstrated in Fig. \ref{fig:result3}(E), such small value of the parameter broadens the bistability regime of the galactose network, causing the right limit point to occur at higher $G_i$ concentrations and making the monostable regime of the induced steady state less attainable.

As in the previous protocol, the GAL2-mutant model is again subjected to a background medium of 0.2 \% w/v galactose and 0.125 \% w/v glucose for 24 hours (to allow the system to reach equilibrium), followed by the addition of an oscillatory glucose input signal of various amplitudes and periods. The responses of the mutant strain according to the 5D and 9D models are shown in \ref{app:ftl}, Fig. B1, in terms of the concentration of Gal1 protein and GAL1 mRNA, respectively. Qualitatively, in both cases, the output signals show an adaptation behaviour similar to the numerical results seen for the wild-type strain models of panels B and C in Fig. \ref{fig:result6}. At low frequencies, however, the mutant model shows no elevation in baseline, unlike the wild-type model.  

For a thorough understanding of the results presented in Fig. \ref{fig:result6}, we use several measures to characterize the oscillatory output signal $S_i$ (where $i=1,\cdots,6$ is the total number of input signals) generated numerically at steady state when both the input and output signals are oscillating in tandem with each other. Four measures, defined in Table \ref{tab:properties}, have been used; these include the normalized mean ($\mathcal{M}_i$), the normalized amplitude ($A_i$) and the upstroke phase ($S_i$) of the signal as well as the phase difference, or phase shift ($\phi_i$) between the phase of the input and output signals, as defined by the Hilbert transform. Given that \citet{bennett2008} varied the frequency of the glucose input signal, we apply here a similar strategy, by calculating these measures across a whole range of frequencies.

\begin{table}\hspace{-0.5cm}
\small
\begin{tabular} {cl} \hline
\vspace{-6pt}\\
\textbf{Measures}&\textbf{Notation and definitions}\\
\vspace{-6pt}\\  \hline \vspace{-6pt}\\
Normalized & $\mathcal{M}_i=\frac{\bar{S_i}}{\max\limits_{i=1\cdots N}(\bar{S_i})}$, where ${\overline{S}}_i=\frac{1}{L}\int\limits_0^L{\text{output }}dt$, $L=20$ min and $N=6$ is the\\
mean & total number of input signals tested. \vspace{6pt}\\
Normalized& $A_i=\frac{\left(\max\limits_{[0,L]}(S_i)-\min\limits_{[0,L]}(S_i) \right)/2}{\max\limits_i\left\{\left(\max\limits_{[0,L]}(S_i)-\min\limits_{[0,L]}(S_i) \right)/2\right\}}$,   where $S_i$ is the output signal for all $i=1,\cdots,6$.  \\
amplitude&\vspace{6pt}\\
Upstroke phase&$U_i$: The duration of the upstroke between a maximum and a preceding minimum\\
& averaged over the period $T$ of the output signal $S_i$, $i=1,...,6$. \\
Phase shift&$\phi_i=\phi_{\rm{input}_i}-\phi_{\rm{output}_i}:$ The difference between the phase of the input signal ($\phi_{\rm{input}_i}$)\\ 
&and the output signal ($\phi_{\rm{output}_i}), i=1,\cdots,6$, as determined by the Hilbert transform\\
&defined in \ref{app:A}. \\ \vspace{-6pt}\\ \hline
\end{tabular}
\caption[Properties of oscillatory output signals.]{The four measures used to characterize the output signals of the 5D and 9D models.}
\label{tab:properties}
\end{table}

The numerical results associated with these four measures are plotted in \ref{app:ftl}, Fig. B.2, and explicitly listed in Table \ref{tab:results5} for the wild-type and the GAL2 mutant, as defined by the 5D and 9D models. More specifically, Fig. B.2 shows that an increase in the frequency leads to a low decrease in the baseline and to a prominent decrease in the amplitude of the output signals $G_1$ and $M_1$ for the 5D and the 9D models, respectively. The figure also shows that there is little variation between the reported results for the two model strains, on the order of 10$^{-5}$ for the baseline and the phase difference  and 10$^{-6}$ for the normalized amplitude (see Table \ref{tab:results5}). One of these results is qualitatively consistent with those of \citet{bennett2008} who showed that the GAL network can low-pass filter glucose, the repressor of the network, by decreasing both the amplitude and the phase while increasing the frequency. Although the models presented here can produce one aspect of this low-pass filtering capacity (namely, the decrease in amplitude), they cannot produce the decrease in the phase shift at high frequencies (see Fig. B.2(B) of \ref{app:ftl}). Indeed, our simulations show that at high frequencies, the system responds rapidly to glucose and peaks earlier when responding to low frequencies.

A possible source for this discrepancy between our results and the reported experimental data is the method employed for calculating the phase difference; \citet{bennett2008} used recorded inputs to calculate phase shifts, which often appear to drift upward and exhibit a decrease in amplitude. These two issues may, as a result, affect the peaks of the input and the overall phase shift values (reported to vary between 0 and $-3\pi/2$ in experimental settings). For our simulation, we used a pure sinusoidal for the glucose repressor oscillations and calculated the input and output phases using Hilbert transform (as shown in Table \ref{tab:properties}). According to such an input signal, the phase shift occurs between [-$\pi$, -$\pi/2$]  (see Table \ref{tab:results5}), and no decrease in the phase difference is observed, as stated earlier. We do observe, however, similar results when using the same pure sinusoidal input signal applied to the model of \citet{bennett2008}. These results suggest that the galactose network does not filter out repressor fluctuations of high frequencies but rather adapts by oscillating with a low amplitude.

To assess the similarity of the output signal to the pure sinusoidal input signal used in our simulation, we measure the upstroke and the downstroke fractions of the cycle, as defined in Table \ref{tab:properties}. Table \ref{tab:results5} and Fig. B.2(C) (in \ref{app:ftl}) show that although the wild-type and the GAL2 mutant strains, defined by the 5D and 9D models, exhibit similar characteristics, the 5D model produces a stable 55:45 ratio between the downstroke and the upstroke phases of the cycle for all frequencies, but the 9D model gradually shifts this ratio from 1:1 to 55:45 when increasing the frequency. Although the two models eventually reach the same ratio at high frequencies, the adaptive behaviour of the 9D model at low frequency is likely to be due to the presence of slow rates in the model providing it with more time to adapt to an idealized sinusoidal signal. 

\begin{table}\hspace{-0.5cm}
\small
\begin{tabular} {ccccccccc} \hline
\vspace{-6pt}\\
\textbf{Measures}&\textbf{Model}&\textbf{Yeast}& \textbf{Period}& \textbf{of the}&\textbf{Input}& \textbf{Signal}&\textbf{(bold)}&\textbf{[hr]} \\
& & \textbf{Type}&\textbf{4.5} &\textbf{3.0} & \textbf{2.25}& \textbf{1.5} & \textbf{1.125} & \textbf{0.75}\\
\vspace{-6pt}\\ \hline
Normalized&5D&WT/GAL2$\Delta$ & 1.00 &  1.00  &  0.99 &  0.99& 0.99&  0.99\\ 
Baseline &9D&WT/GAL2$\Delta$ & 1.00&	0.99&	0.98&	0.97&0.97&0.96\\ \hline
Normalized&5D&WT/GAL2$\Delta$ &1.00&0.68&0.51&0.34&0.26&0.17\\
Amplitude&9D&WT/GAL2$\Delta$ & 1.00&0.90&0.80&	0.63&0.51&0.36\\ \hline
Phase difference &5D& WT/GAL2$\Delta$ &  -1.80&-1.74&-1.68&-1.65&	-1.64&	-1.62 \\
($\phi_i$) [rad]& 9D&WT/GAL2$\Delta$&-2.69&-2.53&-2.38&-2.17&	-2.05&-	1.90\\ \hline
Upstroke percentage& 5D&WT& 45.01&	44.83&	44.818&	44.778&	44.74&44.73\\
of oscillations [\%]&5D& GAL2$\Delta$&  45.02&	44.83&44.81&	44.67&	44.74&	44.67\\ 
& 9D&WT& 49.07&	48.06&	47.26&	46.22&	45.63	&45.33\\
&9D& GAL2$\Delta$ & 49.07&	48.06&	47.33&	46.33&	45.63&45.33\\ \hline
\end{tabular}
\caption[The properties of the 5D and the 9D models, for both the wild type and the mutant yeast strains.]{The properties of the 5D and the 9D models (with $n=1$), for both the wild type (WT) and the GAL2 mutant yeast strains. The values for the baselines and amplitudes of the oscillations are the same within the first two decimal places. The phase difference between the input-output signals (glucose-$G_1$ for the 5D model, and glucose-$M_1$ for the 9D model) also showed a striking similarity between the two types of strain. The upstroke phase of the oscillations occupied a smaller percentage than the downstroke phase, decreasing with increasing frequency for both strains.}
\label{tab:results5}
\end{table}

\subsubsection{Bistability with respect to glucose}
\begin{figure}[!htb]
\hspace{-1cm}
\includegraphics[width=18cm]{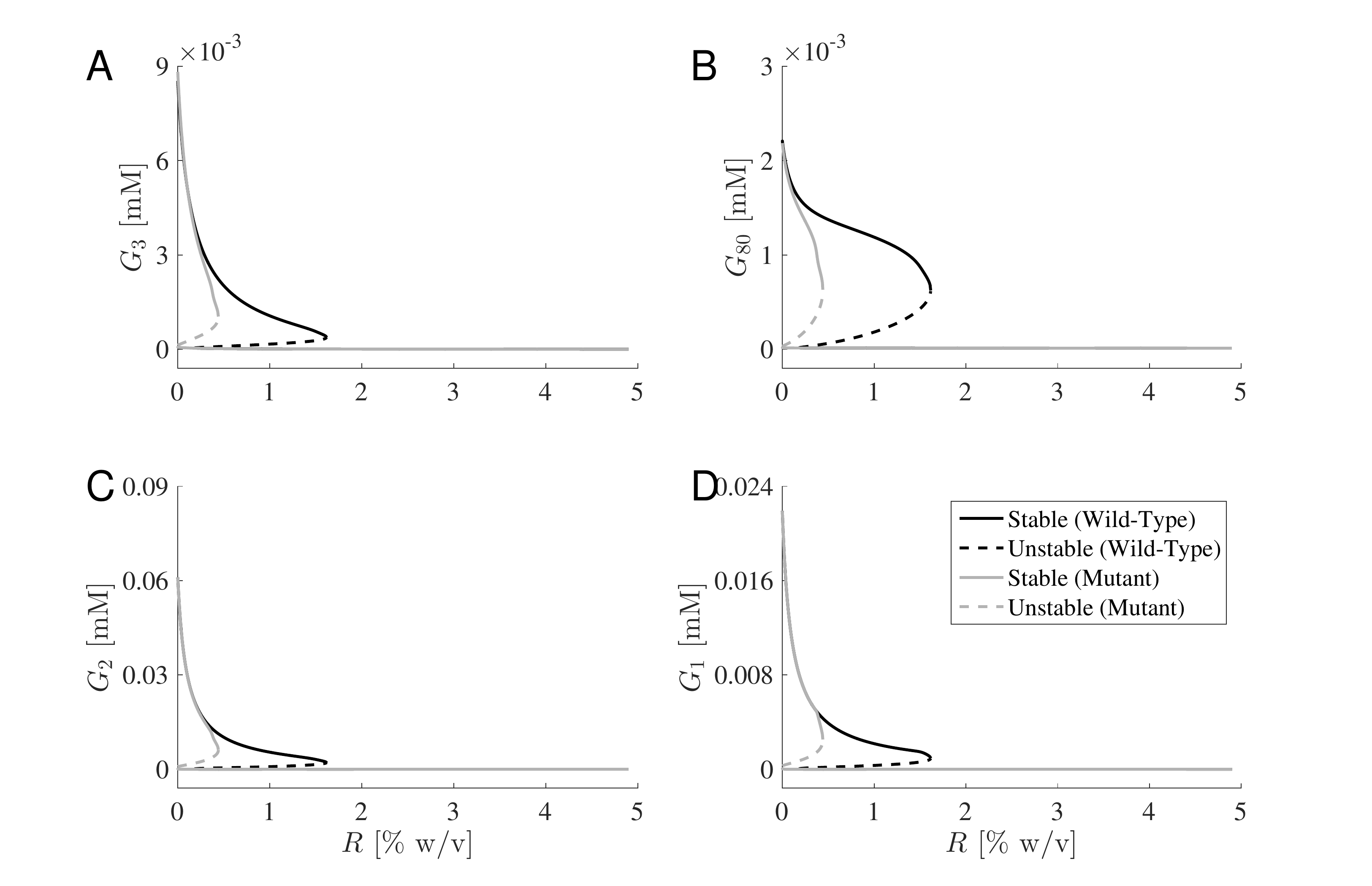}
\caption[One-parameter bifurcation of GAL proteins as a function of glucose]{One-parameter bifurcation of GAL proteins as a function of glucose ($R$), measured in units of [\% w/v]. The four panels show the steady state values of (A) the regulatory protein Gal3 ($G_3$); (B) the inhibitory regulatory protein Gal80 ($G_{80}$); (C) the permease Gal2 ($G_2$); and (D) the regulatory and enzymatic protein Gal1 ($G_1$). Solid lines represent the stable branches of attracting equilibria, whereas dashed lines represent the unstable branches of equilibria, for both wild-type (black) and GAL2 mutant (grey) yeast strains.}
\label{fig:result9}
\end{figure}

\noindent
As suggested in the Section \ref{subsec:glu}, yeast cells show optimal growth on glucose, by acting as a repressor of the GAL network through four independent processes:  (i) by increasing cellular growth rate (or, equivalently, by increasing the dilution rate $\mu$), (ii) by enhancing vesicle degradation of $G_2$ transporter, (iii) by repressing the transcription of GAL3, GAL1 and GAL4, and (iv) by competing with galactose to bind with $G_2$. These processes have been all included in the 9D model described by Eqs. (\ref{eqs:model-glu9}a-i). To analyze the relation between the dynamical properties of the model and the oscillatory input signals, we plot the bifurcation diagrams of the various Gal proteins with respect to glucose for both the wild-type and the GAL2 mutant strains (Fig. \ref{fig:result9}). Bistability in the expression level of $G_3$ (panel A), $G_{80}$ (panel B), $G_2$ (panel C) and $G_1$ (panel D) is observed in all cases at low values of glucose, whereas monostability (determined by the uninduced state) is only observed in one regime at intermediate to high values of glucose. Interestingly, these panels show that although a decrease in the induced (upper) stable branch is observed during an increase in glucose, they remain slightly more elevated than the non-induced (lower) branch at the limit point. The switch from the non-induced to the induced stable branch at the limit point is consistent with that seen experimentally in the level of GAL1 induction \citep{bennett2008}. 

\begin{figure}[!htb]
\centering
\includegraphics[width=10cm]{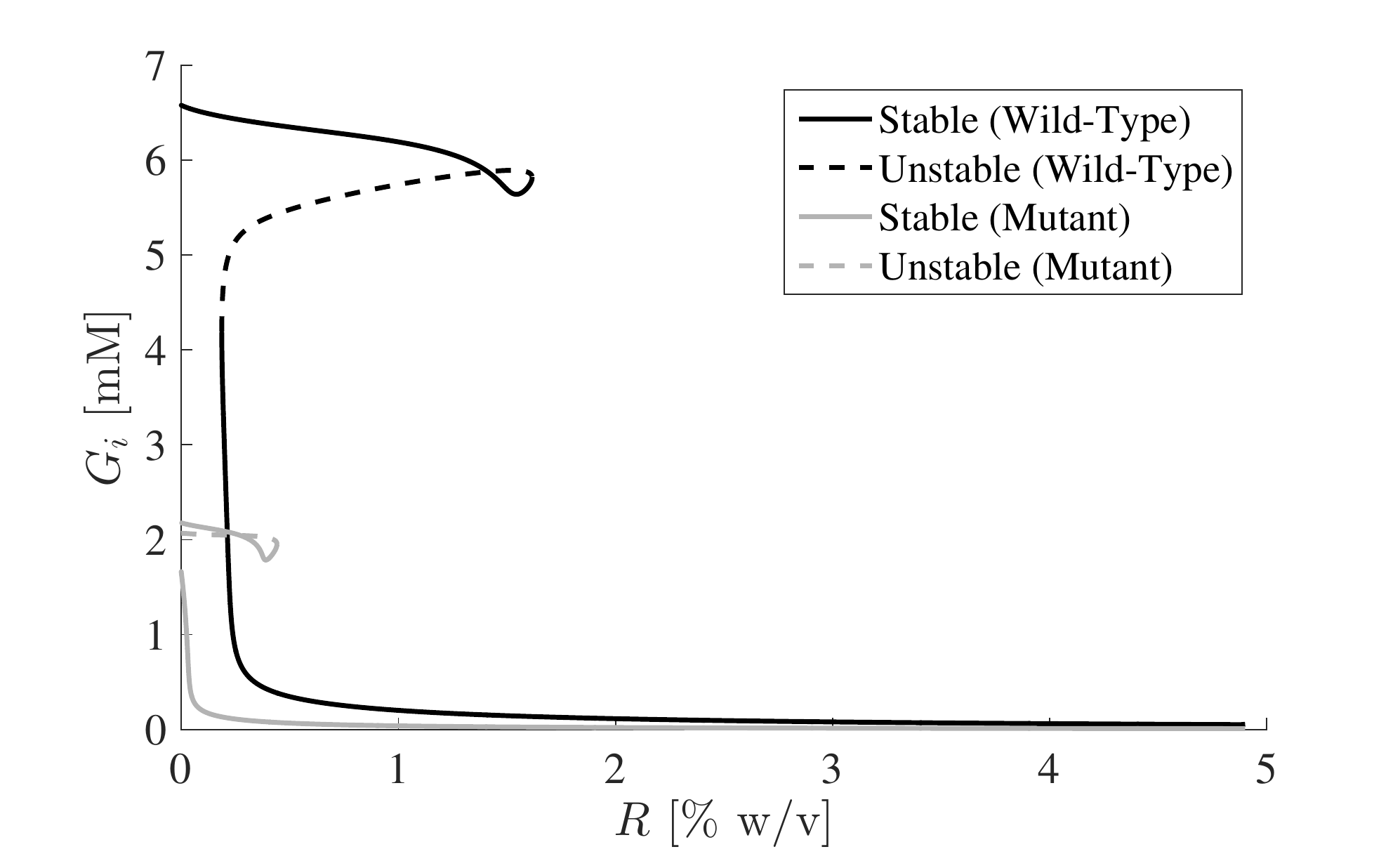}
\caption[One-parameter bifurcation of intracellular galactose concentration ($G_i$) with respect to glucose]{One-parameter bifurcation of intracellular galactose concentration ($G_i$) with respect to glucose ($R$), for both wild-type (black) and GAL2 mutant (grey) yeast strains. As before, solid and dashed lines define the stable and unstable branches, respectively.}
\label{fig:result10}
\end{figure}

Plotting the bifurcation diagram of intracellular galactose ($G_i$) with respect to glucose in Fig. \ref{fig:result10}, we observe a fold around the right limit point situated at about 1.8 \% w/v glucose. This feature is likely due to the multidimensionality of the system. Unlike the bistability of Fig. \ref{fig:result2}, Fig. \ref{fig:result10} shows a large difference between the induced state and the uninduced state inside the bistable regime, an outcome that should be testable experimentally. It is important to point out though that the proximity of the unstable branch to the stable branch of uninduced steady states makes the basin of attraction of the steady states in that branch small. Therefore, to test for bistability, initial concentrations of the various proteins of the GAL network must be chosen carefully. The steep increase in $G_i$ seen in the uninduced state inside the bistable regime is analogous to the sharp increase in the Gal protein expression, especially $G_2$, seen in the induced states of Fig. \ref{fig:result9}. 

Recall that the GAL2 mutant strain is simulated by decreasing the transport rate $\alpha$. This causes the induced stable branch to shift downward for all GAL components, as seen in Fig. \ref{fig:result9}. The shape of the bifurcation diagram in Fig. \ref{fig:result10} can be used to explain how narrowing down of the bistability regime affects the output response to glucose periodic forcing. Given that the noninduced stable branches are similar for both strains, we expect the system in both cases to tend to the noninduced stable branch, if the initial conditions are low. Therefore, narrowing down the bistability regime by decreasing $\alpha$ does not affect the properties of the oscillations during glucose periodic forcing because of the presence of peaks in the input signal (causing significant repression in the expression level of the GAL proteins) and the ability of the network to adapt quickly. An additional prediction drawn from our model is that if glucose oscillatory input signal is varied between 0 and 3\% w/v, the galactose network would cross the right limit point of the bifurcation diagrams shown in Figs. \ref{fig:result9} and \ref{fig:result10} and as a result shift back and forth between the bistable and monostable regimes. 

\section{Discussion} \label{discussion}
\noindent
In this study, we have shown that the bistability property of the GAL regulon, originally seen in a model of \cite{apostumackey} is still preserved in an improved model of the full galactose network. This new model considers additional regulatory and metabolic processes, including (i) the dimerization of regulatory proteins, (ii) the existence of multiple upstream activating sequences at the level of the promoter and (iii) the metabolic reactions involving galactose and glucose repression processes. The agreement between the newly developed model and that of \citet{apostumackey} regarding bistability is due to the fact that the two models assume GAL3 and GAL80 are induced only three times more in galactose as compared to raffinose \citep{ideker2001}, whereas GAL1 and GAL2 are respectively induced 800 times and 1000 times more \citep{sellickreece2008}. In our model, these two assumptions make the metabolic pathways exhibit a prominent switch-like phenomenon in the form of bistability in the Gal proteins.

Although dimerization and multiple upstream activating sequences have been already previously examined in \citet{atauri2004}, bistability was not taken into consideration in that study. This latter feature was actually investigated by other groups \citep{acar2005,acar2010,venturelli2012,venturelli2015} that developed models with simplified feedback loops and nonlinear functions of Michaelis-Menten type to produce results consistent with those observed experimentally. In this study, we combined all of these features in our modeling approach and used the \textit{in vivo} results of \citet{abramczyk2012}, that showed the localization of the tripartite complexes acting as transcription factors (outside and inside the nucleus), to decipher the dynamics of the GAL network. We included both the short and the long-term complexes in our modeling construct. 

The models developed here contained a minimalistic branch representing the metabolic reactions to understand their interaction with the regulatory gene network. We used several numerical techniques to estimate the values of the kinetic parameters of the reactions involved, particularly those induced by glucose administration, such as transcriptional regulation, the competition for the Gal2  transporter and its degradation. Based on these estimations, we were able to reproduce the results obtained experimentally, including the ones associated with the response of wild type and GAL2 mutant strains to periodic forcing by glucose, suggesting that the model is accurate physiologically. Based on these models we were able to show that this network is adaptable to changing environment (such as oscillatory signals) like other metabolic networks that exhibit robustness to nonhomeostatic conditions \citep{nijhout2014}. The models also revealed that large discrepancies between responses of different strains or cells can 
be generated if the transport rate $\alpha$ is decreased to a lower level than its default value or if the external glucose and galactose conditions are made more extreme than those applied experimentally.

It is important to point out that in the models developed here, we did not include the regulatory protein Gal4 and the metabolic enzymes Gal7p and Gal10p for various reasons. First, as it has been already mentioned in \cite{apostumackey}, the expression level of Gal4 protein is not affected when cells are transferred from raffinose to galactose medium \citep{sellickreece2008}. As for the metabolites generated due to the downstream enzymes Gal7p and Gal10p, they do not feed back into the gene network and their respective negative feedback processes are represented through the phosphorylation rate of galactose by the Gal1p kinase. Limiting the number of dynamical equations to the ones involving the key proteins has allowed us to gain a better understanding of how the different molecular constituents interact at the level of the regulon.

The main goal of this study is to determine how the feedback loops of the whole network interact together to form emergent behaviour in various experimental conditions. That includes defining its response to the inducer and the repressor (i.e. galactose and glucose), as well as its adaptability to changing environment (i.e., robustness to nonhomeostatic conditions). Using bifurcation analysis, we demonstrated that the bistability switch, an intrinsic property of the GAL regulon, persists in the full model and plays an integral part in the dynamics of the network. It is very pronounced by most variables (except for intracellular galactose) and underlies many of the features observed experimentally (including the binary response). 

There are several predictions made in this study that can be further explored experimentally. For example, the models presented here showed that intracellular galactose concentration depends on the extracellular conditions giving rise to an induction curve similar to the one presented in Fig. \ref{fig:result2}. This result can be interpreted as the transport mechanism being the main rate limiting step in GAL network induction. However, galactose also participates in other metabolic reactions  that are responsible for its conversion into galactitol, galactonate and other by-products. We expect all these metabolic processes to happen in tandem and the relation between the extracellular and intracellular galactose concentrations to follow the qualitative behaviour predicted by our numerical simulations.

From a clinical perspective, one idea presented here was whether bistability could play a role in galactosemia. Potential causes for this disease have been already revealed, mostly linked to genetic mutations that cause accumulation of galactitol in various tissues. From a mathematical point of view, one can study this disease via parameter perturbations that can lead to an increase in the activation of some reverse rates and a decrease in the accumulation of harmful metabolites. One interesting prediction of our model is that the half-maximum activation of the transport ($K$) is the main parameter that controls the minimal galactose concentration required for bistability (see Fig. \ref{fig:result3}(F)). By translating the bistable regime to larger concentrations of extracellular galactose (i.e., to the right of the current bifurcation diagram), greater values of $K$ would reduce the likelihood of the organism to be fully induced at small galactose concentrations. Altering this constant experimentally, 
perhaps by blocking or modifying the transporter, would decrease the amount of galactitol, the toxic metabolite. A natural continuation of the present work would be to try to describe a threshold for the toxic levels of galactose and its other metabolites, in the context of this disease, particularly galactosemia of type III induced by impairments in the Gal7 enzyme and connected with accumulation of cellular Gal1P \citep{gitzelmann1995}.

Another property of the system deduced from our mathematical models is the interplay between the repressor and the inducer of the galactose network. The bifurcation diagrams plotted with respect to glucose (Figs. \ref{fig:result9}(A-D) and \ref{fig:result10}) show that a high glucose level impedes galactose accumulation, which in turn decreases Gal1P level.  An experimental protocol similar to the ones employed in \citet{brink2009} or \citet{acar2005} can verify our predictions, by measuring the levels of Gal1P for different combinations of galactose and glucose concentrations. Although the two monosaccharides (glucose and galactose) are processed differently, a combination of the two pathways can be beneficial for the yeast cells whose metabolic machinery is not properly functional.  

The lack of experimental data has made our modeling effort a challenging one, relying mostly on estimation techniques. To further validate the models against experimental data, our assumptions on transcriptional repression and competition for the Gal2p transporter due to glucose concentration should be tested experimentally. In the models, we have described the effects of the repressor by Michaelis-Menten type functions whose parameters were estimated using the Genetic Algorithm (see \ref{app:A}). To assess our predictions, experiments can follow the procedures of \citet{barros1999}, to measure sugar transport, and of \citet{lashkari1997}, to obtain mRNA fold-difference values in cells grown in different galactose and glucose mixtures. 

The complexity of this network and its obscurity makes the use of mathematical models an alternative and a promising tool to decipher its kinetics. The rapid discovery of new pathways adds more emphasis on the importance of using such modeling approaches to accomplish this goal. The incorporation of new pathways into the models will allow us to study their effects, including their role in stabilizing/destabilizing its induced and uninduced steady states and in defining its adaptability to environmental perturbations. It will also allow us to predict emergent phenomena exhibited by the model and determine its effects on the physiology of the network. These steps must be accompanied by model validation against experimental data to make sure that conclusions reached are reasonable. These steps were followed very closely when developing the two (5D and 9D) models. 

\section*{Acknowledgments}
\noindent
This work was supported by grants to Anmar Khadra and Michael C. Mackey from the Natural Sciences and Engineering Research Council (NSERC).

\appendix
\setcounter{table}{0}
\renewcommand{\thetable}{A.\arabic{table}}
\setcounter{figure}{0}
\renewcommand{\thefigure}{A.\arabic{figure}}

\section{Parameter Estimation and Software} \label{app:A}

\subsection{Measured Parameters}
\begin{itemize}
\item Conversion constant ($c$)

\noindent
Although there is a large variation in the shape of yeast cells and their volume we consider the generic yeast cell to be spherical, haploid cell, of volume $70$  $\mu$m$^3$. This value has been published in \cite{sherman2002} and has been used in other computational models \citep{ramsey2006,apostumackey}.
 
 The cellular volume and Avogadro's number are two constants required for estimating $c$, defined to be the conversion constant in Table \ref{table:par-estim}. This parameter is included to maintain the consistency of units between [mM] and [molecules/cell], and is given by:
 \begin{subequations}
 \begin{align*}
 1 \text{ mM}&= \frac{10^{-3} \text{ mol} }{1 \text{ L}}= \frac{ 6.02214129\times 10^{23}\times10^{-3} \text{ molecules}}{(1 \text{ dm})^3}=\\
 &=\frac{ 6.02214129\times 10^{20} \text{ molecules}}{(10^{-5}\text{ } \mu \text{m})^3}\times \frac{70\text{ } \mu \text{m}^3}{\text{cell}}=4.2154989 \times 10^7 \text{ molecules/cell}=c.
 \end{align*}
 \end{subequations}
\item {Dilution rate ($\mu$)}

 \noindent
 The dilution rate is often calculated by using the doubling time of the cells. \cite{ramsey2006} and \cite{apostumackey} used a doubling time of 180 mins in their models, which is equivalent to 3.85 $\times 10^{-3} \text{ min}^{-1}$. However, to obtain consistency between our parameter choices for modelling glucose repression, we averaged of the doubling rates reported by \cite{tyson1979}, to obtain $156$ min in galactose and $79$ min in glucose. Based on the above, we conclude that
 \begin{subequations}
 \begin{align*}
  &\mu_a=\frac{\ln 2}{156 \text{ min}}\approx 4.438 \times 10^{-3} \text{ min}^{-1}\\
  &\mu(R)_{max}=\mu_a+\mu_b=\frac{\ln 2}{79 \text{ min}}\approx 5.12 \times 10^{-3} \text{ min}^{-1}.
 \end{align*}
 \end{subequations}
\item {mRNA degradation rate  $(\gamma_{M})$}

\noindent
This parameter has been measured experimentally as the half life-time of the mRNA strands. \cite{wang2002} measured this quantity for an average strand, whereas \cite{bennett2008} measured it for GAL3 and GAL1. We have approximated the half-life time at 16 min, which is equivalent to
 \begin{equation*}
  \gamma_M=\frac{\ln 2}{16 \text{ min}}\approx 43.32\times 10^{-3} \text{ min}^{-1}.
 \end{equation*}
\item {Protein degradation rates  $(\gamma_{G,i}, i\in\{3,80,2,1\})$}

\noindent
The values were initially measured in
\cite{holstege1998} and \cite{wang2002} and used in the model developed by  \citet{ramsey2006}. Here, we use the same parameters as in  \citet{ramsey2006}, except that the protein degradation in our model does not represent the degradation arising from cellular growth, but rather from protein processing only. This implies that
\begin{subequations}
 \begin{align*}
 \gamma_{G,3}&= 11.55\times 10^{-3} \text{ min}^{-1}- 4.438\times 10^{-3} \text{ min}^{-1}=7.112\times 10^{-3} \text{ min}^{-1},\\
 \gamma_{G,80}&= 6.931\times 10^{-3} \text{ min}^{-1}- 4.438\times 10^{-3} \text{ min}^{-1}=2.493\times 10^{-3} \text{ min}^{-1},\\
 \gamma_{G,2}&\approx 0 \text{ min}^{-1},
 \gamma_{G,1}\approx 0 \text{ min}^{-1}.
 \end{align*}
\end{subequations}
\item {Transcription rates ($\kappa_{r,3},$ $\kappa_{r,80},$ $\kappa_{r,2},$ and $\kappa_{r,1}$)}

\noindent
Transcription rates are estimated using similar approach to that presented by \cite{apostumackey}. More specifically, they are approximated by using mRNA steady state ratios, measured in cells grown in induced versus repressed extracellular media. In other words, their numerical values ($\kappa_{r,i}$, $i=3, 80, 2, 1$) are calculated by setting Eqs. (6), (8), (10) and (12a), describing the dynamics of mRNA, to 0, and solving for $\kappa_{r,i}$ in terms of the steady state values $M_{i,(ss)}$ as follows:
$$\kappa_{r,i}=M_{i,(ss)}\frac{\gamma_{M}+\mu_a}{100 \%},\quad i\in\{3,80,2,1\}.$$

The mRNA steady state levels in glucose were estimated in \cite{arava2003} to be 0.8, 1.1 and 1.0 molecules/cell for GAL3, GAL80 and GAL1, respectively. Since there are no estimates available for the steady state level of Gal2p, we used instead the steady states of mRNAs for all hexose transporters reported by Arava and colleagues. Table 3 in the appendix of \cite{arava2003} contains the copy numbers for 17 hexose transporters. We use all these values, except for one that is particularly high (5 compared to a range of [0.2, 2.6] for the other hexose copy numbers values) to calculate the median, which gives the approximate value of 0.598 mRNA copies/cell.

To calculate the transcription rates in galactose, we also consider the fold difference between mRNA values in galactose-grown compared to glucose-grown cells, as reported by \cite{lashkari1997}. Based on this premise, we have
\begin{subequations}
 \begin{align*}
 &\kappa_{r,i}=\text{mRNA}_{\text{level in glucose}}\times\text{fold number} \times{(\gamma_M+\mu_a)}\\
 &\kappa_{r,3}= 0.8 \text{ molecules/cell} \times 8.6  \times (4.78 \times 10^{-2} \text{ min}^{-1})=0.329 \text{ molecules/(cell} \times \text{min)}\\
 &\kappa_{r,2}=  0.598 \text{ molecules/cell} \times 23.7  \times (4.78 \times 10^{-2} \text{ min}^{-1})=0.678 \text{ molecules/(cell} \times \text{min)}\\
 &\kappa_{r,1}=  1.0 \text{ molecules/cell} \times 21.8  \times (4.78 \times 10^{-2} \text{ min}^{-1})=1.042 \text{ molecules/(cell} \times \text{min)}
  \end{align*}
\end{subequations}
\begin{subequations}
 \begin{align*}
\text{Upper bound of }\kappa_{r,80}&=1.1 \text{ molecules/cell} \times 3.0 \times (4.78 \times 10^{-2} \text{ min}^{-1})\\
&=0.158 \text{ molecules/(cell} \times \text{min)}\\
\text{Lower bound of }\kappa_{r,80}&=1.1 \text{ molecules/cell} \times 2.8 \times (4.78 \times 10^{-2} \text{ min}^{-1})\\
&=0.147 \text{ molecules/(cell} \times \text{min)}.\\
 \end{align*}
\end{subequations}
\item {Translation rates ($\kappa_{l,3},$ $\kappa_{l,80},$ $\kappa_{l,2},$ and $\kappa_{l,1}$)}

\noindent
The four parameters associated with the transcription rates of GAL3, GAL80, GAL2 and GAL1 are not measured experimentally using the desired units and the sugar medium that we require for the model. \cite{arava2003} presented in Table 3 of their supplementary information the protein synthesis rates in glucose media for an extensive list of mRNA strands, in units of [proteins/sec]. 
Since we are interested purely in the translation rates $\kappa_{l,i}$  ($i=3, 80, 2, 1$), in units of [proteins/mRNA copies $\times$ cell], we estimate these parameters based on the following equation
 \begin{equation} \label{eq:kli}
\kappa_{l,i}=\frac{\text{fraction of translated mRNA} \times \text{elongation rate}}{\text{protein length}}\times \frac{\text{number of proteins}}{\text{mRNA}},  
 \end{equation}
where the ``fraction of translated mRNA'' is assumed here to be equivalent to the ``relative translation rate'', defined in \cite{arava2003} to be:
 $$\text{relative translation rate}=\text{ribosome occupancy} \times \text{ribosome density}.$$
 
 For this relation, the ribosome occupancy is approximated by the ribosomal mRNA level divided by the total mRNA level of that species available in the cell, and the ribosome density is given by the number of ribosomes per length of unit of the open reading frame. The relative translation rate, calculated in this manner in \cite{arava2003}, is therefore unitless and is given by 0.143, 0.039 and 0.042 for GAL3, GAL80 and GAL1, respectively. For GAL2, the lack of experimental data constrains us to use the median relative translation rates of all associated GAL mRNAs reported, which is given by 0.145.

 As for value of the mRNA elongation rate, it was reported in \cite{arava2003} to be 10 amino acids (a.a.) per second, for the yeast cells grown in YPD medium (i.e., 1\% yeast extract, 2\% peptone and 2\% dextrose). This rate is similar to that obtained by \cite{bonven1979}, who found that is about 9.3 a.a per second for budding yeast grown in glucose instead of a mix of peptone and dextrose. 

 To calculate the translation rates of Eq. (\ref{eq:kli}), we still need the length of the protein, measured in number of amino acids, and the ratio of proteins to mRNA. We already know that Gal proteins measure: 520, 435, 528, and 574 amino acids (for Gal3p, Gal80p, Gal1p and Gal2p, respectively). Furthermore, \cite{ideker2001} found that the ratios of proteins to mRNA lies between 4200 and 4800. Based on the above observations, we can now calculate the translation rates of Eq. (\ref{eq:kli}) as follows
 
 \noindent
 Lower bounds:
\begin{subequations}
 \begin{align*}
 &\kappa_{l,3}=\frac{0.143\times 9.3 \times 60 \text{ a.a./min}}{520 \text{ a.a}}\times 4200\frac{\text{proteins}}{\text{mRNA}}=644.49 \frac{\text{proteins}}{\text{mRNA}\times\text{min}}.\\
&\kappa_{l,80}=\frac{0.039\times 9.3 \times 60 \text{ a.a./min}}{435 \text{ a.a}}\times 4200\frac{\text{proteins}}{\text{mRNA}}=210.12 \frac{\text{proteins}}{\text{mRNA}\times\text{min}}.\\
&\kappa_{l,1}=\frac{0.042\times 9.3 \times 60 \text{ a.a./min}}{528 \text{ a.a}}\times 4200\frac{\text{proteins}}{\text{mRNA}}=186.42 \frac{\text{proteins}}{\text{mRNA}\times\text{min}}.\\
&\kappa_{l,2}=\frac{0.145\times 9.3 \times 60 \text{ a.a./min}}{574 \text{ a.a}}\times 4200\frac{\text{proteins}}{\text{mRNA}}=592.02 \frac{\text{proteins}}{\text{mRNA}\times\text{min}}.
\end{align*}
\end{subequations}
 We have used these lower bounds for the translation rates, as shown in Table \ref{table:par-estim}.
\item {Dimerization constants}

\noindent
\cite{melcher2001} reported that the dimerization constant for $G_{80,d}$ ($K_{D,80}$) is 1 to 3$\times 10^{-7}$ mM. In our model simulations, we used the aforementioned upper bound. 
\item {Dissociation constants ($K_{B,80}$ and $K_{B,3}$)}

\noindent
The dissociation constant of the Gal80p dimer from the promoter conformation $D_2$ is measured to be 3 $\times 10^{-8}$ mM,  in \cite{melcher2001}, and 5$\times 10^{-6}$ mM, in \cite{lohr1995}. As for the dissociation constant of the activated Gal3p from the Gal80p molecule, it was numerically estimated by \cite{venkatesh1999} to be 6$\times 10^{-8}$ mM. We use this latter numerical value to estimate the dissociation constant $K_{B,3}$. This value represents the rate of a single $G_{80}$ binding to an activated $G_3$ molecule in the cytoplasm, in the context of a different kind of GAL model (i.e., based on nucleo-cytoplasmic shuttling of $G_{80}$). In our model, however, we consider the reaction between dimers of each of these species, not single molecules. Due to the lack of relevant data we still use these values as an approximation and set 
$$K_{B,80}=5\times 10^{-6} \text{ mM},\quad K_{B,3}=6\times 10^{-8} \text{ mM}.$$
\item {Transport rate ($\alpha$)}

\noindent
For the transport rate, we use, as a reference frame, the parameter value 4350 min$^{-1}$ provided in \cite{atauri2005}. The authors of this latter study mention that the rate was adjusted in order to obtain a $V_{max}$ consistent with that observed experimentally in \cite{reifenberg1997}.
\item {Parameters involved in galactose phosphorylation ($\kappa_{GK}, K_m, K_{IC}, K_{IU}$)}

\noindent
The rate parameters associated with phosphorylation have been previously estimated in the experimental paper of \cite{timsonreece2002}. No other manipulations and calculations are necessary, since the units and the definitions of all rate constants are in agreement with the ones used in our models (see Table \ref{table:par-estim}).
\item {Metabolic rate ($\delta$)}

\noindent
A suitable candidate for estimating this parameter is the rate of the reaction catalyzed by the Gal7p transferase enzyme, which ensures that Gal-1-Phosphate is metabolized and incorporated into the glycolytic pathway. This catabolic rate of the transferase ($k_{cat,GT}$) has been measured as $59200$ min$^{-1}$ and used in the study of \cite{atauri2005}. Thus we set
 $$\delta \approx k_{cat,GT}=59200 \text{ min}^{-1}.$$
 \end{itemize}
\subsection{Parameters estimated through the model}
\begin{itemize}
\item {Half-maximum activation constant of $G_3$ and $G_1$ ($K_{S}$)}

\noindent
It has been known for some time that galactose induces the entire regulatory system via the activation of Gal3p molecule, but the actual reactions involved in this induction process remain incompletely understood. Several modelling papers focusing on this topic have assumed that this process follows either Michaelis-Menten activation kinetics \citep{venkatesh1999}, similar to the formalism used here, or linear kinetics \citep{acar2005,atauri2005}. This process is also modelled in terms of a positive constant added to the rates of change of the different proteins induced by galactose \citep{venturelli2012}.

Given that a Michaelis-Menten formalism has been used to describe activation, the value of half-maximum activation of these reactions is assumed to be identical to the estimated numerical value of 4000 mM, given in \cite{apostumackey}. By having a very large half-maximum activation constant, we are assuming an almost-linear relationship between the concentrations of activated Gal3 and Gal1 proteins and the intracellular galactose, which would be in agreement with other modelling papers that have used a direct proportional (or linear) relationship.
\item {Parameters of the regulatory pathways involving Gal3p and Gal1p ($k_{cat,3}$, $K_{D,3}$, $k_{cat,1}$, $K_{D,1}$ and $K_{B,1}$)}
\noindent
As indicated earlier, the process of Gal3p and Gal1p activation is not fully understood. By using QSS assumption on the model, we can derive relations between the different parameters of the model based on its steady states.

 According to Eq. \ref{eq:Ks}, we have
 \begin{subequations}
\begin{align*}
K_{3}&=\frac{\sqrt{K_{D,3}K_{B,3}}(\gamma_{G,3}+\mu)}{\kappa_{C,3}}\\
 K_{1}&=\frac{\sqrt{K_{D,1}K_{B,3}K_{B,1}}(\gamma_{G,1}+\mu)}{\kappa_{C,1}}.
  \end{align*}
\end{subequations}

Due to the fact that bistability is one of the main properties of the GAL network (observed within a given physiological range of galactose concentration), one can use this property to numerically estimate the two parameters $K_3$ and $K_1$. In other words, the values of these parameters can be determined by ensuring that the model can produce bistability. Based on this, we find that $K_3=1.729 \times 10^{-6}$ mM$^2$ and $K_1=3.329\times 10^{-6}$ mM$^2$. Since we already know the degradation rates, the dilution rates for these proteins as well as the activation rate $K_{B,3}$ , we can solve for the following ratios in term of the parameters $K_{D,i}$, $K_{B,1}$, $\kappa_{C,i}$:
 \begin{subequations}
\begin{align*}
 \frac{\sqrt{K_{D,3}}}{\kappa_{C,3}}&=\frac{K_3}{\sqrt{K_{B,3}}(\gamma_{G,3}+\mu)}=4.363
\times 10^{-3} \text{ mM}^{1.5}\text{min}\\
 \frac{\sqrt{K_{D,1}K_{B,1}}}{\kappa_{C,1}}&=\frac{K_1}{\sqrt{K_{B,3}}(\gamma_{G,1}+\mu)}=3.062 \text{ mM}^{1.5}\text{min}.
 \end{align*}
\end{subequations}

 These ratios are used in the functions $K_3$ and $K_1$:
\begin{subequations}
\begin{align*}
K_{3}&=\sqrt{K_{B,3}}(\gamma_{G,3}+\mu)\times 4.363 \times 10^{-3} \text{ mM}^{1.5}\text{min}\\
K_{1}&=\sqrt{K_{B,3}}(\gamma_{G,1}+\mu)\times 3.062 \text{ mM}^{1.5}\text{min}.
\end{align*}
\end{subequations}
\end{itemize}
\subsection{Numerical estimation of parameters}
\begin{itemize} 
\item {Software}

\noindent
The numerical results that we have presented in the main text were obtained using two main software packages: {\tt XPP} (written by Bard Ermentrout and freely available online), used for numerical bifurcation analysis, and {\tt MATLAB}  (MathWorks Inc., 2014, Natick, Massachusetts), used for simulating the differential equation models. ``Cftool" and ``ga" toolboxes available in {\tt MATLAB} were used to fit various expressions to experimental data as explained below.

\item {Data fitting}
\begin{itemize}
\item Parameter estimation using ``Cftool"

\noindent
As shown in Figure \ref{fig:appB1} and Table \ref{tab:appB1}, the half-maximum activation for the repressive functions characterizing the dilution and Gal2p degradation rates are estimated using Michaelis-Menten functions in the ``Cftool" toolbox (with R = 1, SSE = $5.5\times 10^{-14}$ for the dilution rate, and R = 1, SSE = $1.12\times 10^{-14}$ for the degradation rate).

This toolbox is also used to estimate the phase of the glucose oscillatory inputs administered in the experiments of \cite{bennett2008}. This was done by digitizing the the data presented in the first  rows of panels a and b in Figure
3 of that paper and fitting each outcome to a sinusoidal signal with a given period. The baseline was allowed to vary during this process.

\begin{figure}[!htb] 
\includegraphics[width=16.5cm]{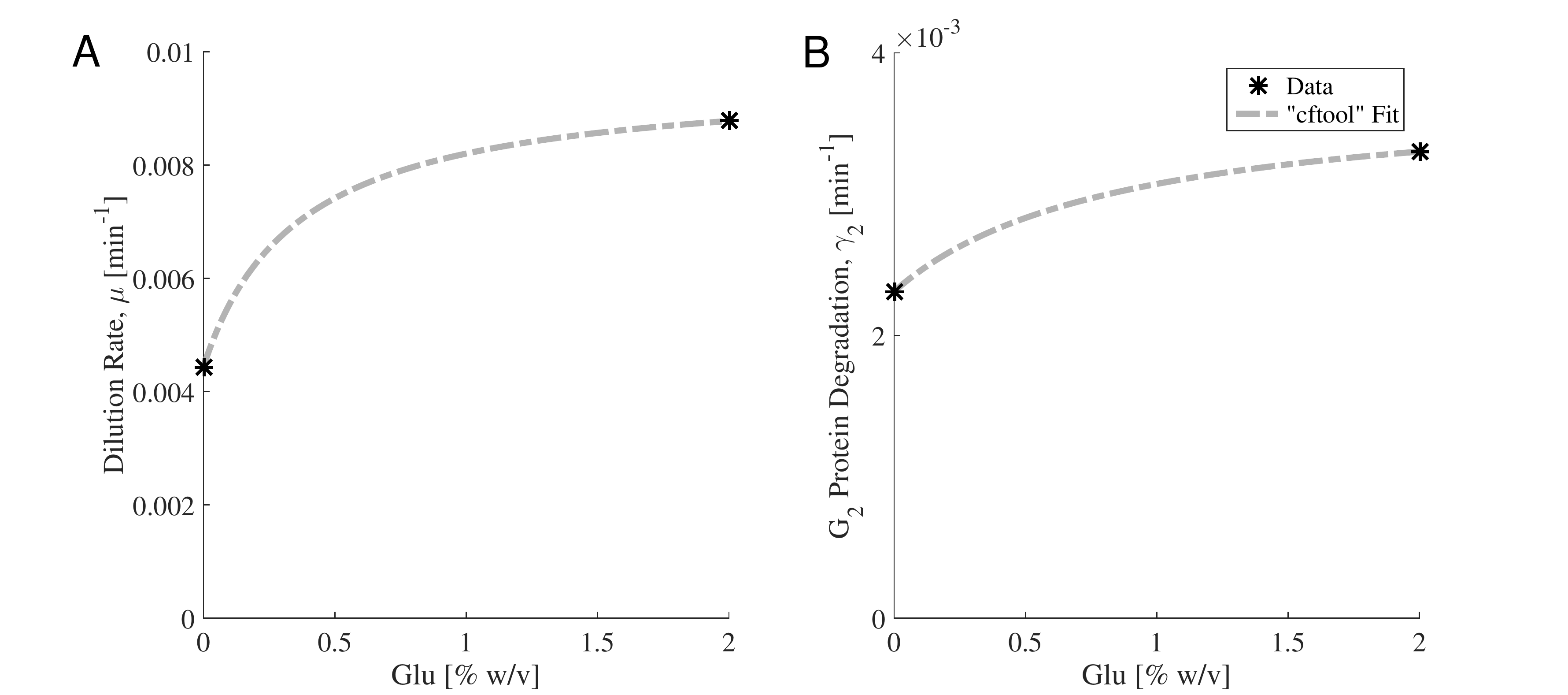}
\caption[Cellular processes affected by glucose: dilution and degradation]{Cellular processes affected by glucose. (A) An increase in glucose concentration leads to an increase in the dilution rate ($\mu$). (B) An increase in extracellular glucose induces an increase in the degradation rate of the transporter protein $G_2$ ($\gamma_2$). Black asterisks represent experimental data obtained from \cite{tyson1979} and \cite{horakwolf}, whereas dashed grey lines represent the fitted Michaelis-Menten functions using the ``Cftool" toolbox in {\tt MATLAB}.}
\label{fig:appB1}
\end{figure}

\begin{table}
\begin{tabular}{c c c c c} \hline
\vspace{-6pt}\\
\textbf{Processes}& \textbf{Expression}&\textbf{Parameters}& \textbf{Value}& \textbf{Reference}\\
\vspace{-6pt}\\ \hline
Dilution & $\mu=\mu_a+\frac{\mu_b\times R}{\mu_c+R}$ & $\mu_a$ & $4.44\times 10^{-3} $
& \citep{tyson1979} \\
 &  & $\mu_a+\mu_b$ & $8.78\times 10^{-3} $& \citep{tyson1979}\\
 &  & $\mu_c$ & $5.12 \times 10^{-3} $& Fitted with ``Cftool" \\ \hline
G${}_{2}$ Degradation  & $\gamma_2=\gamma_a\ +\frac{\gamma_b\times R}{\gamma_c+R}$ & $\gamma_a $ &
$3.98\times 10^{-3} $ & \citep{horakwolf}\\
 &  & $\gamma_a+\gamma_b\ $ & $7.66\times 10^{-3}$  & \citep{horakwolf} \\
 &  & $\gamma_c\ $ & $1.416\times 10^{-3}$  & Fitted with ``Cftool" \\ \hline
\end{tabular}
\caption{Parameter values associated with dilution and GAL2 degradation obtained using a combination of parameter estimation and ``Cftool" fitting.}
\label{tab:appB1}
\end{table}
\noindent

\item Parameter estimation using the genetic algorithm

\noindent
The parameters associated with the transcriptional repression and competition for the $G_2$ transporter ($x_b, y_b, y_c$) are estimated using the Genetic Algorithm. This was done by minimizing the error between the experimental data and the steady state values of the extended 5D and 9D models. Rates of transcriptional repression and galactose transport are both modelled as Hill functions with negative coefficients and fitted to data from \cite{bennett2008} (see Fig. \ref{fig:gafit}). Since the type of inhibition in these processes is unknown, the Hill coefficients are left undetermined and allowed to assume only two values: either 1 or 2. The latter follows from the fact that trimer complexes are far less likely to form than monomers and dimers.

\begin{figure}[!htb] \hspace{-1.4cm}
\includegraphics[width=19cm]{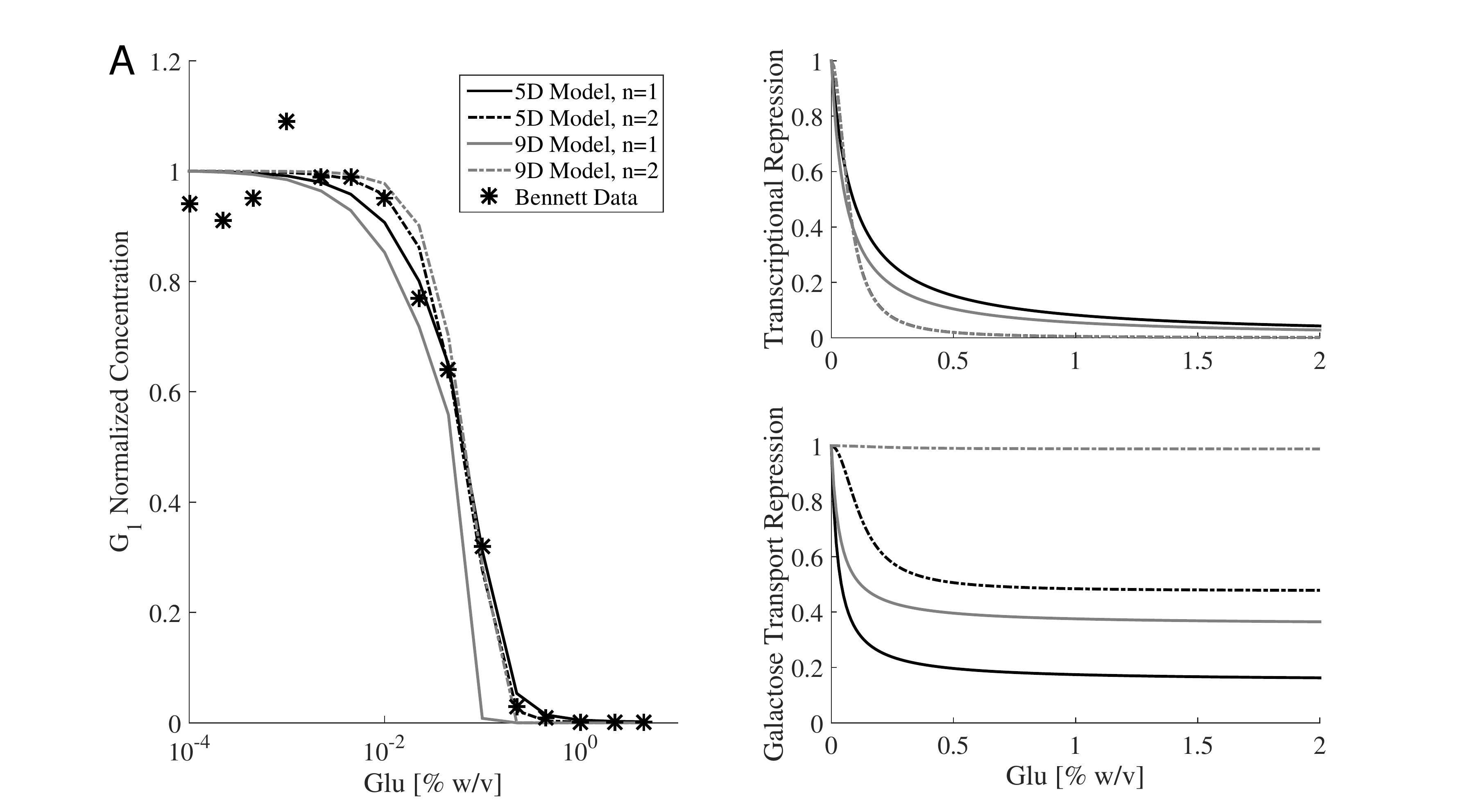}
\caption{Data fitting of glucose-induced repression. (A) Experimental data published by \cite{bennett2008} for the normalized $G_1$ concentration, shown in asterisks, was used to estimate the parameters of the Hill functions for the repressor processes in the extended 5D and 9D models (black and grey lines, respectively). These Hill functions were allowed to have a coefficient of 1 and 2 (solid and dashed lines).  (B) The half-maximum deactivation is similar for the four fitted models. (C) Transport of galactose through the $G_2$ permease is also decreased due to the presence of the repressor. The four models behave differently here, with large variations in the half-maximum deactivation and the asymptotic minimum.}
\label{fig:gafit}
\end{figure}

\begin{table}
\begin{tabular} {l l l l l l } \hline
&  & \textbf{5D} & &
\textbf{9D}& \\ \hline
\textbf{Value of $n$}  &  & 1 & 2 & 1 & 2 \\ \hline
Half-maximum transcriptional repression& $x_c$ & \textbf{0.2443} & 2.4107 & \textbf{0.2657}& 0.0650\\ \hline
Transport competition rate & $y_b$ & \textbf{0.0003} & 0.5639 &\textbf{0.4950} &  0.4950\\ \hline
Half-maximum transport competition& $y_c$ & \textbf{2.9989} & 0.0052& \textbf{2.3142}& 2.3142  \\ \hline
Error estimation &  & \textbf{0.0387} &0.0300&\textbf{0.0876}& 0.1402   \\ \hline
\end{tabular}
\caption{Parameter values associated with the two processes transcriptional repression (by glucose) and competition for the transporter, as determined by the Genetic Algorithm. Parameter combinations that give the best fit are shown in bold.}
\end{table}
\end{itemize}
\item {Hilbert transform}

\noindent
The Hilbert transform allows one to calculate the phase and the amplitude of an oscillatory signal, \textit{u(t)}. In theory, it convolves the signal with the Cauchy kernel, also known as Cauchy Principal value (\textit{p.v.}), given by
$$ p.v.(f(x))=\lim _{a\to \infty } \int _{-a}^{a}f(x)dx,$$
where \textit{f(x)} is a function with the properties
$$ \int_{-\infty }^{0}f(x)dx=\pm \infty \text{ and }\int _{0}^{\infty }f(x)dx=\mp\infty.$$
This type of improper integral is used in the calculation of the Hilbert transform defined by
\begin{equation*}
H(u(t))=\frac{1}{\pi } p.v.\int _{-\infty }^{\infty }\frac{u(\tau)}{(t-\tau )} d\tau.
\end{equation*}
One of the properties of this transform is that it shifts the signal within the integral by $\pi $/2, i.e. $H(H(u))(t)=-u(t).$ It is also related to the Fourier transform by the relation
$$ F(H(t))=F\left(\frac{1}{\pi t} \right)F(u(t)).$$

In discrete form, the Hilbert transform can be calculated in {\tt MATLAB}, using the function `Hilbert' . This function computes the transform by generating a complex signal $H(u(t))=A(t)e^{i\phi(t)}$ from the original signal $u(t)$.  The complex signal is then used to evaluate its amplitude (\textit{A}) and phase (\textit{$\varphi $}) based on the equations
\begin{subequations}
\begin{align*}
A(t)&=\mathcal{R} e(H(u(t))) \\
\varphi (t)&=angle(H(u(t))), \\
\end{align*}
\end{subequations}
where $\mathcal{R}e$ is the real part of the Hilbert transform and \textit{angle} is a predefined {\tt MATLAB} function for the instantaneous phase of the transform within ($-\pi ,\pi $). This method was also used in \cite{khadra2009} to analyze synchrony in a population of synchronized neurons.

As an example of the Hilbert transform technique of calculating the phase of a signal, we will refer to Fig. \ref{fig:hilbert} shown below, containing the estimated phase difference between the oscillatory glucose input signal and $G_1$ output response for the wild-type strain, using the GAL extended 5D model, with $n=1$. This calculation involves applying the Hilbert Transform on both the glucose input and the Gal1p output signals. Notice how the output signal (black lines) follows the input (grey lines), but the lines are not always straight, due to the fact that the output is not purely sinusoidal. 

\begin{figure}[!htb]
\hspace{-1cm}
\includegraphics[width=19cm]{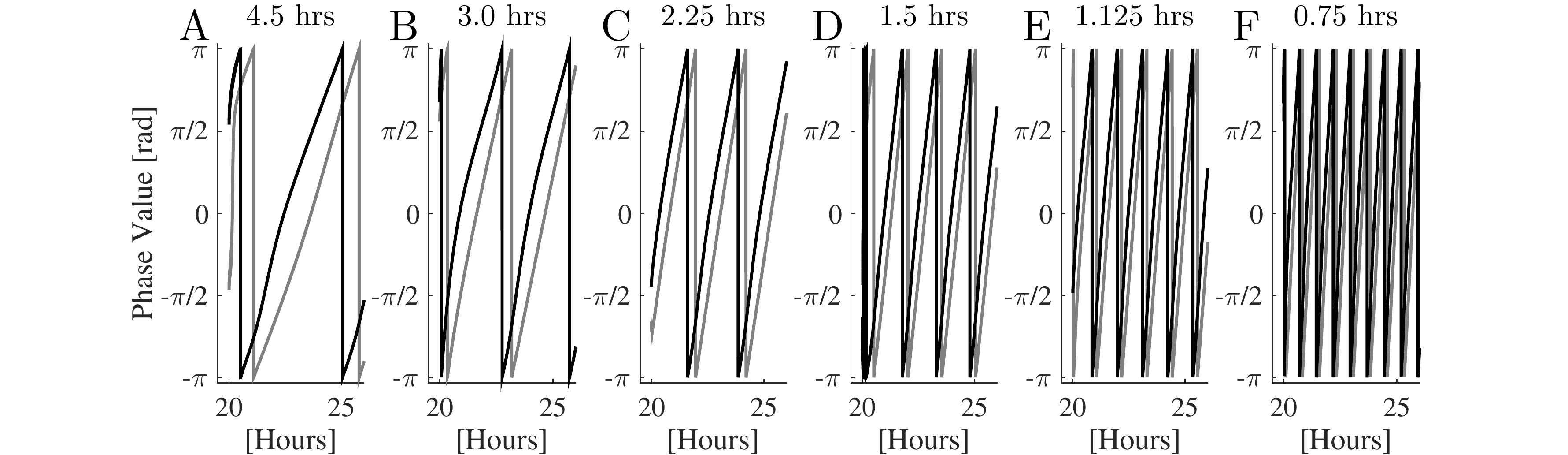}
\caption[Hilbert transform of the oscillatory glucose input and the $G_1$ output signals]{Hilbert transform of the oscillatory glucose input and the $G_1$ output signals, showing the periodicity of both signals and their phase characteristics. (A-F) Glucose (grey lines) and the galactokinase $G_1$ (black lines) oscillatory signals at steady state (i.e., after 20 hours) exhibiting at least one cycle. The output signal follows the input signal very closely and adapts to its periodicity given by (A) 4.5, (B) 3, (C) 2.25, (D) 1.5, (E) 1.125 and (F) 0.75 hours.}
\label{fig:hilbert}
\end{figure}
\end{itemize}

\newpage
\section{Additional Derivations and Results} \label{app:ftl}
\begin{itemize}
\item {Fractional Transcriptional Level} 

\noindent
Here, we describe the complete derivation of the function $\mathcal{R}_n$ presented in Eq. \ref{eq:Rss}. Assuming that the dimerization reactions presented in Table \ref{table:table_rxn} are at equilibrium, we can conclude that the dissociation constants of all relevant molecular species are given by the ratio $\frac{\text{Reactants}}{\text{Products}}$. Based on this, we can derive the following expressions for the dimer molecules $G_{80,d}$, $G^*_{3,d}$ and $G^*_{1,d}$
\begin{subequations}
\begin{align}
K_{D,80}&=\frac{G_{80}^2}{G_{80,d}} \Rightarrow G_{80,d}=\frac{G_{80}^2}{K_{D,80}}\\
K_{D,3}&=\frac{(G^*_{3})^2}{G^*_{3,d}} \Rightarrow G^*_{3,d}=\frac{(G^*_{3})^2}{K_{D,3}}\\
K_{D,1}&=\frac{(G^*_{1})^2}{G^*_{1,d}} \Rightarrow G^*_{1,d}=\frac{(G^*_{1})^2}{K_{D,1}}.
\end{align}
\label{eqs:dc}
\end{subequations}

It is also mentioned in Table \ref{table:table_rxn} that the three dimers $G_{80,d}$, $G^*_3$ and $G^*_1$ have high affinities for the three promoter conformations, $D_2$, $D_3$ and $D_4$, respectively. Assuming that these quantities reach equilibrium quickly, we can write the reactions in terms of their dissociation constants. Using this result and Eqs. (\ref{eqs:dc}a-c), we can express the promoter conformations $D_1$, $D_3$ and $D_4$ in terms of $D_2$ as follows

\begin{subequations}
\begin{align}
K_{B,80}=\frac{G_{80,d}D_1}{D_2} &\Rightarrow D_1=\frac{K_{B,80} D_2}{G_{80,d}}=\frac{K_{D,80} K_{B,80} D_2}{G_{80}^2}\\
K_{B,3}=\frac{G^*_{3,d}D_2}{D_3} &\Rightarrow D_3=\frac{G^*_{3,d}D_2}{K_{B,3}}=\frac{(G^*_{3})^2D_2}{K_{D,3}K_{B,3}}\\
K_{B,1}=\frac{G^*_{1,d}D_3}{G^*_{3,d}D_4} &\Rightarrow D_4=\frac{G^*_{1,d}D_3}{G^*_{3,d}K_{B,1}} =\frac{G^*_{1,d}G^*_{3,d}D_2}{G^*_{3,d}K_{B,1}K_{B,3}}= \frac{G^*_{1,d}D_2}{K_{B,1}K_{B,3}}.
 \end{align}
 \label{eqs:kb}
\end{subequations}
These expressions are then used to formulate $\mathcal{R}_1$, the fractional transcriptional level for mRNA strains containing a single [UAS]$_g$,
$$\mathcal{R}_1=\frac{D_1+D_3+D_4}{D_1+D_2+D_3+D_4}=1-\frac{D_2}{D_1+D_2+D_3+D_4},$$
or equivalently
$$\mathcal{R}_1(G_{80} ,G_{3}^*,G_{1}^*)=1-\frac{1}{1+\frac{K_{D,80}K_{B,80}}{G_{80}^2}+\frac{
(G_3^*)^2}{K_{D,3}K_{B,3}}+\frac{(G_1^*)^2}{K_{D,1}K_{B,1}K_{B,3}}}.$$

As mentioned in the paper, GAL2 and GAL1 mRNAs contain 2 and 5 [UAS]$_g$ sequences respectively. Thus, the promoter can exist in more conformations than the four terms presented above. In this context, we simply need to multiply the kinetic reactions shown in Table 1 by the number of existing [UAS]$_g$ ($n$).  This means that $\mathcal{R}_n$ is given by:
$${\mathcal{R} _{n} (G_{80} ,G_{3}^*,G_{1}^*)}{=1-\frac{1}{1+\sum\limits_{k=1}^n\Big(\frac{\sqrt{K_{D,80}K_{B,80}}}{G_{80}}\Big)^{2k}
+\sum\limits_{k=1}^{n}\Big(\frac{G_{3}^*}{\sqrt{K_{D,3}K_{B,3}}}\Big)^{2k} + \sum\limits
_{k=1}^{n}\Big(\frac{G_{1}^*}{\sqrt{K_{D,1}K_{B,1}K_{B,3}}}\Big)^{2k}}.}$$

\item {Quasi steady state of the Galactose-1-Phosphate}

\noindent
Here, we present the derivation of the rate of change for intracellular galactose levels ($G_i$). This is done by applying QSS assumption on the rate of change of $G_p$ (given by Eq. \ref{eq:gp}) to obtain 
\begin{equation*}
(\delta+\mu_a)G_{p(ss)}^2+(\delta+\mu_a)k_p (G_i)G_{p(ss)} -\sigma(G_i ) G_i G_1=0,
\end{equation*}
where $G_{p,ss}$ represents the steady state value of Gal1P ($G_p$). Solving for $G_{p,ss}$ gives one single positive solution
\begin{subequations}
\begin{align}
&G_{p(ss)} =\frac{-k_p (G_i )+\sqrt{k_p (G_i )^2+\frac{4\sigma(G_i)G_iG_1}{\delta}}}{2}.
\end{align}
\end{subequations}
The expression for $G_{p(ss)}$ can be substituted into Eq. \ref{eq:gi} for the intracellular galactose.
\newpage
\item {Dynamics of the model simulation depicting the response of GAL2$\Delta$ strain to glucose oscillations}
\begin{figure}[!htb]
\hspace{-1cm}
\includegraphics[width=19cm]{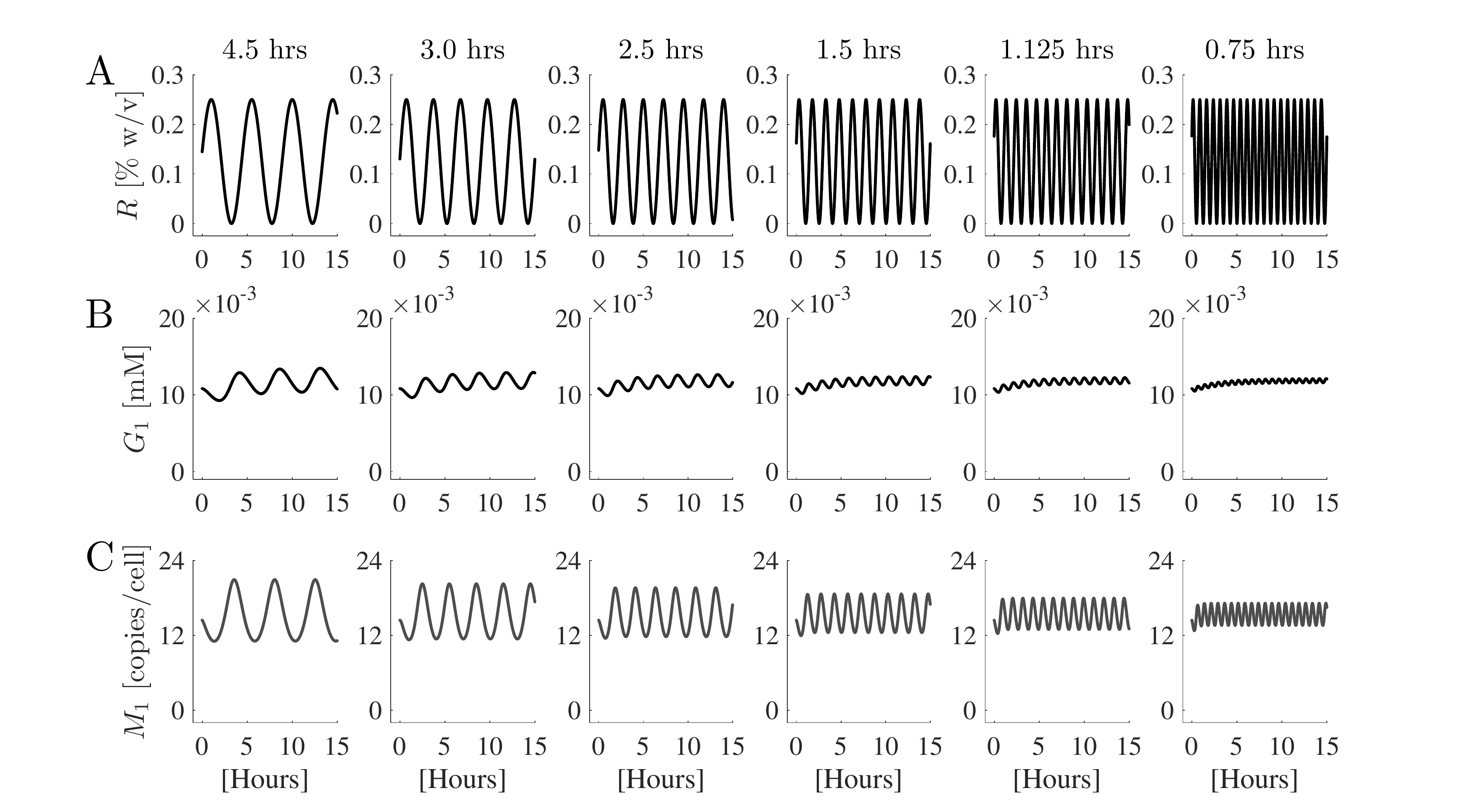}
\caption{Model response to oscillatory glucose input signal, generated using a GAL2 mutant cell (A) The oscillatory glucose input signal applied with a period defined on top of each panel. (B) Gal1 output signal, generated from the extended 5D model, showing a similar 5 hours period as that seen in Fig. \ref{fig:result9}(B). (C) GAL1 mRNA output signal, generated from the 9D model, showing a slightly lower baseline when compared to the wild-type strain of Fig. \ref{fig:result9}(C). }
\label{fig:result7}
\end{figure}
\item {Quantification of the four measures defining the output response of the GAL network}
\begin{figure}[!htb]
\hspace{-2cm}
\includegraphics[width=19cm]{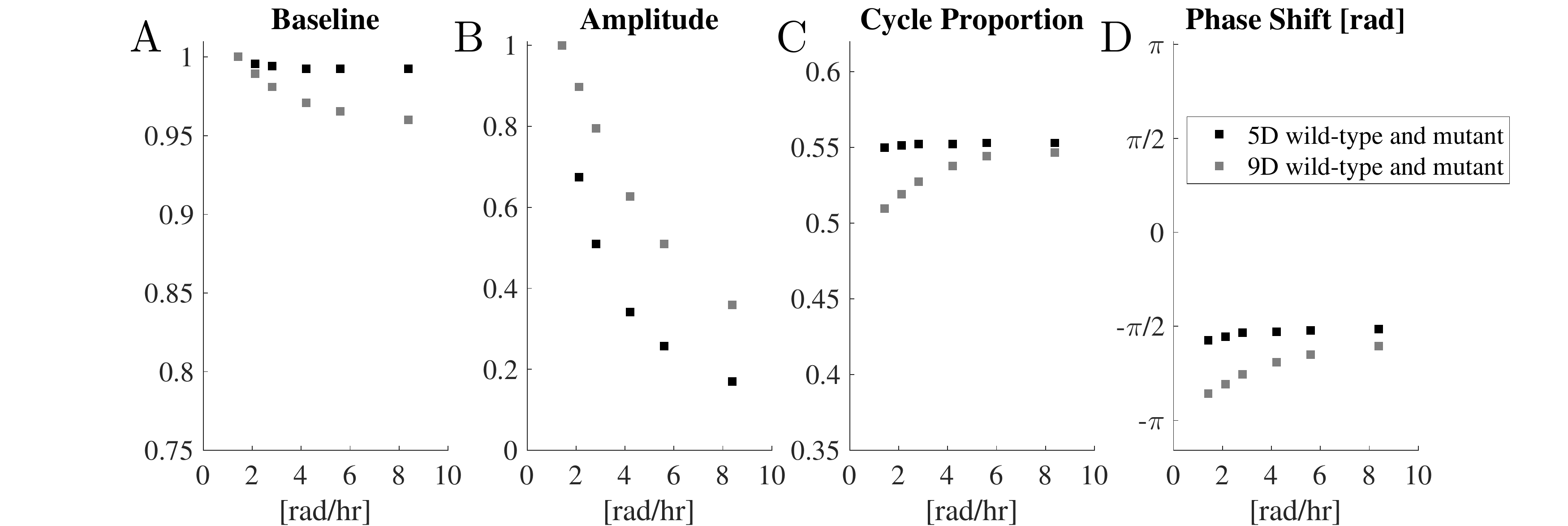}
\caption{Comparison between the properties of the $G_1$ oscillatory output signal in wild-type and GAL2 mutant strains, as determined by the numerical simulations of the 5D and 9D models in Figs. \ref{fig:result6} and \ref{fig:result7}. The four characteristic measures defined in Table \ref{tab:properties} are used; namely, the (A) baseline, (A) amplitude, (C) percentage of the upstroke, and (D) phase shift. Notice the minimal change in the phase shift.}
\label{fig:result8}
\end{figure}
\end{itemize}
\newpage

\section*{References}
\bibliographystyle{elsarticle-harv}
\bibliography{GalactoseLibrary}

\end{document}